\documentclass[twocolumn,abs,prb,showpacs]{revtex4-1}

\usepackage{graphicx}
\usepackage{dcolumn}
\usepackage{float}
\usepackage{amsmath}
\usepackage{bm}
\usepackage[mathscr]{euscript}

\newcommand{\comment}[1]{}

\begin{document}


\title{Optical and Transport Properties in 3D Dirac and Weyl Semimetals}

\author{C. J. Tabert$^{1,2}$}
\author{J. P. Carbotte$^{3,4}$}
\author{E. J. Nicol$^{1,2}$}
\affiliation{$^1$Department of Physics, University of Guelph,
Guelph, Ontario, Canada N1G 2W1} 
\affiliation{$^2$Guelph-Waterloo Physics Institute, University of Guelph, Guelph, Ontario, Canada N1G 2W1}
\affiliation{$^3$Department of Physics, McMaster University,
Hamilton, Ontario, Canada L8S 4M1} 
\affiliation{$^4$Canadian Institute for Advanced Research, Toronto, Ontario, Canada M5G 1Z8}
\date{\today}

\begin{abstract}
{Within a Kubo formalism, we study dc transport and ac optical properties of 3D Dirac and Weyl semimetals.  Emphasis is placed on the approach to charge neutrality and on the differences between Dirac and Weyl materials.  At charge neutrality, the zero-temperature limit of the dc conductivity is not universal and also depends on the residual scattering model employed.  However, the Lorenz number $L$ retains its usual value $L_0$.  With increasing temperature, the Wiedemann-Franz law is violated.  At high temperatures, $L$ exhibits a new plateau at a value dependent on the details of the scattering rate.  Such details can also appear in the optical conductivity, both in the Drude response and interband background.  In the clean limit, the interband background is linear in photon energy and always extrapolates to the origin.  This background can be shifted to the right through the introduction of a massless gap.  In this case, the extrapolation can cut the axis at a finite photon energy as is observed in some experiments.  It is also of interest to differentiate between the two types of Weyl semimetals: those with broken time-reversal symmetry and those with broken spatial-inversion symmetry.  We show that, while the former will follow the same behaviour as the 3D Dirac semimetals, for the zero magnetic field properties discussed here, the latter type will show a double step in the optical conductivity at finite doping and a single absorption edge at charge neutrality.  The Drude conductivity is always finite in this case, even at charge neutrality. 
}
\end{abstract}


\maketitle

\section{Introduction}

The last ten years have seen a remarkably rapid development of an entirely new 
direction in condensed matter physics, that of so-called Dirac materials and 
the related area of topological insulators. The origins of this novel field reach back to 1947 when Wallace provided the low-energy dispersion of two-dimensional (2D) graphene\cite{Wallace:1947}.  Decades later, it was the seminal work of Semenoff\cite{Semenoff:1984} and others which 
recognized that the 2D honeycomb lattice could provide a solid state analogue 
for 2D massless Dirac fermions.  This provided the possibility for a condensed matter manifestation 
of phenomena previously discussed in the context of high energy physics.\cite{Neto:2009, Abergel:2010}  The
subsequent isolation of graphene\cite{Novoselov:2004,Novoselov:2005a} and explosion of experimental results\cite{Neto:2009, Abergel:2010} 
verifying its exotic phenomena, have proved one of the most exciting developments of
the last decade. Such success has emboldened scientists to seek further 
variations on the theme of Dirac materials. Initial investigations focused
on 2D systems, proposing adaptations such as strain, buckling, etc. and expanding to
bilayers and other configurations. 

A parallel development, also based on the honeycomb lattice, was the 
suggestion by Kane and Mele of a 2D topological insulator (TI) phase\cite{Kane:2005a,Kane:2005} 
induced by a spin-orbit energy gap which could give rise to a quantum spin 
Hall effect. Insulating in the bulk, these materials harbour conducting edge 
states. Other researchers furthered these ideas to examine the effect of 
spin-orbit coupling in three-dimensional (3D) materials to produce 3D TIs\cite{Hasan:2010,Qi:2011,Ando:2013}. 
These also house insulating bulk states but with Dirac-like surface states\cite{Hasan:2010,Qi:2011,Ando:2013}. This led to the 
discovery of a host of materials providing evidence for these new 
phases of matter. Key to this development has been the power of
density functional theory (DFT) to provide predictions of materials that would
manifest TI and Dirac states of matter (see, for example, Refs.~\cite{Wang:2012a,Wang:2013a,Weng:2014,Weng:2015, Huang:2015a}). The success of these methods, coupled with the experimental realization of these 
materials and confirmation of the Dirac physics, has given rise to an
overwhelming amount of research in a very short time.  

With the success of the graphene field and the 3D TI development with its
attendant 2D Dirac surface state, it became natural to contemplate the
possibility of a 3D analogue to the 2D Dirac system. This implies a dispersion
which is linear at low energy in all three $k$-space directions with particle
and hole states above and below a node. The term 3D Dirac semimetal (DSM) is 
now being used to describe such a state and, near the Dirac point, it is
characterized by a 4x4 Hamiltonian\cite{Lundgren:2014,Klier:2015}. For the DSM, there exists a twofold 
degeneracy in the conduction and valence bands. This degeneracy can be lifted
in a crystal which is noncentrosymmetric (spatial-inversion symmetry
breaking) or has time-reversal symmetry breaking facilitated by magnetism, 
for example\cite{Hosur:2013, Burkov:2015a}. In the former, the degenerate Dirac cone is split into two
cones that are shifted in energy relative to each other. In the latter case, the
two cones are shifted in $k$-space. The two nondegenerate 
cones are referred to as Weyl cones with Weyl points due to their low-energy behaviour 
mapping to a 2x2 Weyl Hamiltonian instead of a Dirac Hamiltonian. As they result from the splitting of a Dirac cone, Weyl cones always come in pairs of nondegenerate cones of opposite chirality\cite{Wan:2011}. This
type of material is termed a Weyl semimetal (WSM). While the bulk WSM has a low-energy linear dispersion about the Weyl point, the surface of the WSM is predicted
to have Fermi arcs whose tips are linked, through projection, to the two bulk
Weyl points of opposite chirality. These materials are expected to demonstrate
exotic properties such as large negative magnetoresistance in the
presence of parallel electric and magnetic fields due to the chiral anomaly\cite{Nielsen:1983,Gorbar:2014}.
Hence, there has been considerable excitement and intense research to find
real materials which demonstrate the behaviour of 3D WSMs and also 3D DSMs.

Over this last year, experimental realizations of 3D DSMs have been announced 
in Cd$_3$As$_2$\cite{Neupane:2014,Borisenko:2014,Jeon:2014,Liu:2014a} and Na$_3$Bi\cite{Liu:2014,Xu:2013}, with other related possibilities discussed in
quasicrystals\cite{Timusk:2013}, HgCdTe\cite{Orlita:2014} and other systems\cite{Gibson:2015}. WSMs
have now been reported in TaAs\cite{Lv:2015,Xu:2015a,Lv:2015b,Xu:2015d,Yang:2015a} and NbAs\cite{Xu:2015b} for WSMs with broken spatial-inversion
symmetry and YbMnBi$_2$ for a WSM with broken time-reversal symmetry\cite{Borisenko:2015}. Other
candidate systems for WSMs might be TI heterostructures\cite{Burkov:2011a}, pyrochlore irridates\cite{Wan:2011},
HgCr$_2$Se$_4$\cite{Xu:2011a}, etc. Indeed, angle-resolved photoemission spectroscopy (ARPES)
experiments are confirming the presence of the linear dispersions and the
existence of Fermi arcs. There is also work on the cyclotron orbits that result from these surface state arcs\cite{Potter:2014,Moll:2015} which have spin texture\cite{Lv:2015a}.  Very recently, a flurry of activity has occurred surrounding the negative magnetoresistance expected and observed due to the chiral anomaly\cite{Kim:2013,Li:2014,Huang:2015,Zhang:2015a,Shekhar:2015,Yang:2015,Li:2015,Zhang:2015b,Li:2015a,Liang:2015}.  Clearly, these are exciting developments and the 
field is on the edge of a huge outpouring of new developments in both theory
and experiment. In this context, we consider predictions
for dc transport measurements and finite frequency conductivity in
zero magnetic field with application to both DSMs and WSMs.

In the realm of dc and ac magneto-electrical transport, 
several theoretical works have already appeared, covering
a range of topics from weak to strong disorder, different types of impurity scattering 
including charge impurity scattering, electron screening, and so forth. For further information, the reader is directed to Refs.\cite{Burkov:2015a,Burkov:2011,Hosur:2012, Huang:2013, Son:2013, Ominato:2014,Gorbar:2014,Kobayashi:2014, Burkov:2014, Ominato:2015, Rodionov:2015, Burkov:2015}.   Here, we only study the case of zero magnetic field and follow the lead of Lundgren \emph{et al.}\cite{Lundgren:2014} by examining three models for different types of 
impurity scattering to determine their effect on longitudinal electrical transport
properties, considering a single DSM or WSM cone. Specifically, we consider: a constant
scattering rate, the weak scattering (Born) limit, and the case of charge impurities. We
calculate the dc and ac conductivities, the thermal conductivity, the Seebeck coefficient,
and the Lorenz number of the Wiedemann-Franz law. We contrast the results obtained from the
Kubo formula with those of the Boltzmann equation. Most of the literature
has taken the Boltzmann approach; we find that the Kubo formulation
provides an extra term beyond the  Boltzmann result which becomes significant in the limit
of $\mu\to 0$ for $\mu< (\Gamma, T)$, where $\mu$ is the chemical potential measured from
the nodal point, $T$ is the temperature, and $\Gamma$ is the impurity scattering rate.  We
find specific signatures of this extra term in these properties and are able to provide
analytical results for several limits.

The dynamical conductivity is of particular interest given that a small amount of preliminary theoretical work has been done\cite{Burkov:2011a, Hosur:2012,Rosenstein:2013,Ashby:2013,Ashby:2014} emphasizing the linear in photon frequency interband background and the
modification in a magnetic field. Likewise, experiments on various materials
have been observing this linear conductivity in quasicrystals\cite{Timusk:2013}, ZrTe$_5$\cite{Chen:2015}, Kane fermions in Hg-Cd-Te\cite{Orlita:2014},
the pyrochlore irridate Eu$_2$Ir$_2$O$_7$\cite{Sushkov:2015}, and WSM TaAs\cite{Xu:2015}.
This gives impetus to provide a more detailed description of the optical conductivity for DSMs and
WSMs. A feature in some of the data is a non-zero positive-frequency intercept in 
the linear conductivity which is difficult to explain in the simple theory. Consequently,
we consider the case of a massless gapped DSM to illustrate a way of obtaining this
feature in the conductivity. Such a model was proposed by Benfatto \emph{et al.}\cite{Benfatto:2008,Cappelluti:2014} for graphene to
explain some anomalous optical data in the literature\cite{Dawlaty:2008}. More recently, this model has been 
proposed in 3D for a system with a Hubbard $U$ Mott gap\cite{Morimoto:2015}. We find such a model more effective at
explaining the data in 3D DSMs. As a final contribution to the WSM literature, we provide the
optical conductivity that would be expected for a WSM which is formed through the breaking of
spatial-inversion symmetry. In this case, there will be two steps in the linear interband
conductivity, which is distinct from the conductivity of a WSM with time-reversal symmetry breaking,
and therefore would provide a clear signature of which type of WSM one has, i.e., the type with the
cones separated in energy versus that with a separation in momentum.

The structure of our paper is as follows. In Sec.~II.A, we discuss the dc results for the electrical
conductivity, examining the effects of three models for impurity scattering and contrasting
the Kubo formula\cite{Carbotte:2010} result with that from the Boltzmann approach\cite{Qin:2011,Xiao:2006,Sharma:2015,Kim:2014,Son:2012}. Temperature dependence is also included. Subsection B discusses the Lorenz number and the 
Wiedemann-Franz law, along with the Seebeck coefficient $S$. The detailed results for $S$ and the
thermal conductivity $\kappa$ are given in Appendix B, while Appendix A provides the Kubo approach to
the optical conductivity at any frequency. In Sec.~III, we treat the ac conductivity for various
impurity models, including a look at the optical spectral weight variation with temperature. This
section pertains to 3D DSMs and 3D WSMs with time-reversal symmetry breaking. In Sec.~IV, we examine
the differences in the optical conductivity that would occur for the WSM with broken-spatial
inversion symmetry where clear signatures of this state would occur.
Finally, continuing on the theme of optical conductivity, the effect of a massless gap on the
conductivity of a DSM/WSM is presented in Sec.~V, motivated by the experiments which display
deviations from the simple theory. Both the dc and ac conductivity are discussed for a
constant scattering rate. Our final conclusions are found in Sec.~VI.

\section{DC Transport}

In the following, we shall discuss results for a single Dirac cone in 3D described by a low-energy Hamiltonian of the form
\begin{align}
\hat{H}=\hbar v_F \bm{\sigma}\cdot\bm{k},
\end{align}
where $\bm{k}=(k_x,k_y,k_z)$ is the 3D wavevector, $\bm{\sigma}=(\sigma_x,\sigma_y,\sigma_z)$ is the 3D vector of Pauli matrices and $v_F$ is the Fermi velocity.  This Hamiltonian gives the energy dispersion $\varepsilon_{\bm k}=\pm\hbar v_F |\bm{k}|$.  To include more valleys through a valley degeneracy $g_v$ or a spin degeneracy $g_s$, a multiplicative degeneracy factor of $g=g_sg_v$ is all that is needed for modifying the results.  For the properties calculated in this paper, this will apply for both DSMs and WSMs with time-reversal symmetry breaking.  Modification to these results for WSMs with spatial-inversion symmetry breaking will be discussed later as a superposition of the DSM results (see Sec.~IV).

\subsection{Conductivity}

In the one-loop approximation, the real part of the dc conductivity obtained from the Kubo formula [Eqn.~\eqref{dc-app}] is
\begin{align}\label{dc}
\sigma_{\rm dc}=\frac{e^2}{6\pi^2\hbar^3 v_F}\int_{-\infty}^\infty d\omega\left(-\frac{\partial f(\omega)}{\partial\omega}\right)\frac{\omega^2+3\Gamma(\omega)^2}{2\Gamma(\omega)},
\end{align}
where $f(\omega)=[{\rm exp}([\omega-\mu]/T)+1]^{-1}$ is the Fermi function (with the Boltzmann constant $k_B=1$).  This expression applies to any energy-dependent impurity scattering rate model $\Gamma(\omega)$.  For simplicity, we have neglected the corresponding real part of the self-energy which is related to its imaginary part [$\Gamma(\omega)$] through Kramers-Kronig transformation.  At zero temperature, the conductivity reduces to
\begin{align}\label{dc-T0}
\sigma_{\rm dc}=\sigma_0\frac{\mu^2+3\Gamma(\mu)^2}{\Gamma(\mu)},
\end{align}
where $\sigma_0\equiv e^2/(12\pi^2\hbar^3v_F)$ and $\mu$ is the chemical potential which describes charge doping away from the neutrality point.  The first term in Eqn.~\eqref{dc-T0} is recognized as the result one would find via the Boltzmann equation approach. It is proportional to the electronic density of states $N(\omega)$ at the Fermi energy ($\propto \mu^2$ in 3D) and to $1/\Gamma(\mu)$, the scattering time at $\omega=\mu$.  
We stress that this term has the form of the single contribution to 
$\sigma_{dc}$ obtained in the work of Lundgren \emph{et al.} (Ref.~\cite{Lundgren:2014}) based 
on a Boltzmann equation approach to which we wish to compare. Later when 
we consider other DC transport coefficients, we will again compare 
specifically with the results of Ref.~\cite{Lundgren:2014} which we will refer to as the 
Boltzmann term. Returning to Eqn.~\eqref{dc-T0}, the second term goes like the 
scattering rate itself rather than its inverse and is not part of the 
work of Ref.~\cite{Lundgren:2014}. If this second term is neglected and the scattering
rate is interpreted as a transport scattering rate, the remaining term is
identical in both the Kubo and Boltzmann approaches.
Here, we will consider three possible impurity models.  First, we take $\Gamma(\omega)$ to be independent of energy (i.e. $\Gamma(\omega)=\Gamma_0=$ constant).  The second case is the weak-scattering model\cite{Lundgren:2014}
 which can be treated in first-order perturbation theory and gives a scattering rate proportional to the density of states.  That is, $\Gamma(\omega)=\Gamma_1\omega^2$, with constant $\Gamma_1$ in units of one over energy.  Finally, we consider the long-range Coulomb impurity model described in the work of 
Burkov \emph{et al.}\cite{Burkov:2011} 
and used by Lundgren \emph{et al.}\cite{Lundgren:2014} to describe transport in Dirac materials within the Boltzmann equation formulation.  In this case, it is envisaged that the number of charge impurities is equal to the doping ($n$) away from the Dirac point.  Consequently, $\Gamma(\omega)$ is proportional to $n\propto\mu^3$ giving $\Gamma(\omega)=\Gamma_2\mu^3/\omega^2$, where $\Gamma_2$ is a dimensionless constant.

The constant scattering rate model is phenomenological in nature and makes no
assumption about the microscopic origin of the impurity scattering involved. It
is important and is extensively used in discussions of transport and optical
properties as it often captures much of the essential underlying physics.
Nonetheless, as our work will show, the observation of deviations
from this model can provide valuable information on the type of the 
impurity centers involved.

The long range Coulomb impurity model was elaborated upon in the work of 
Burkov \emph{et al.}\cite{Burkov:2011}. 
The idea is that each impurity center is charged with
charge transferred to the semimetal. In this case, which can actually be 
realized in some semimetals, the number of impurity centers and the doping
are the same. This model was further employed in the Boltzmann equation 
approach to the transport properties of Dirac-Weyl semimetals described
in the work of Ref.~\cite{Lundgren:2014} with which we will want to compare. While this specific
model is important in its own right, here it also serves to illustrate how 
it can provide differences from the very simplified constant rate 
phenomenological model as well as from weak scattering in the Born 
approximation. In this last case, there is no compelling
microscopic reason for relating the number of impurities to the number of
dopants. Studying both models independently covers a sufficient
range of possible impurity models to give the reader a good idea of 
how such details can affect transport and optical properties.

\begin{figure}[h!]
\begin{center}
\includegraphics[width=0.90\linewidth]{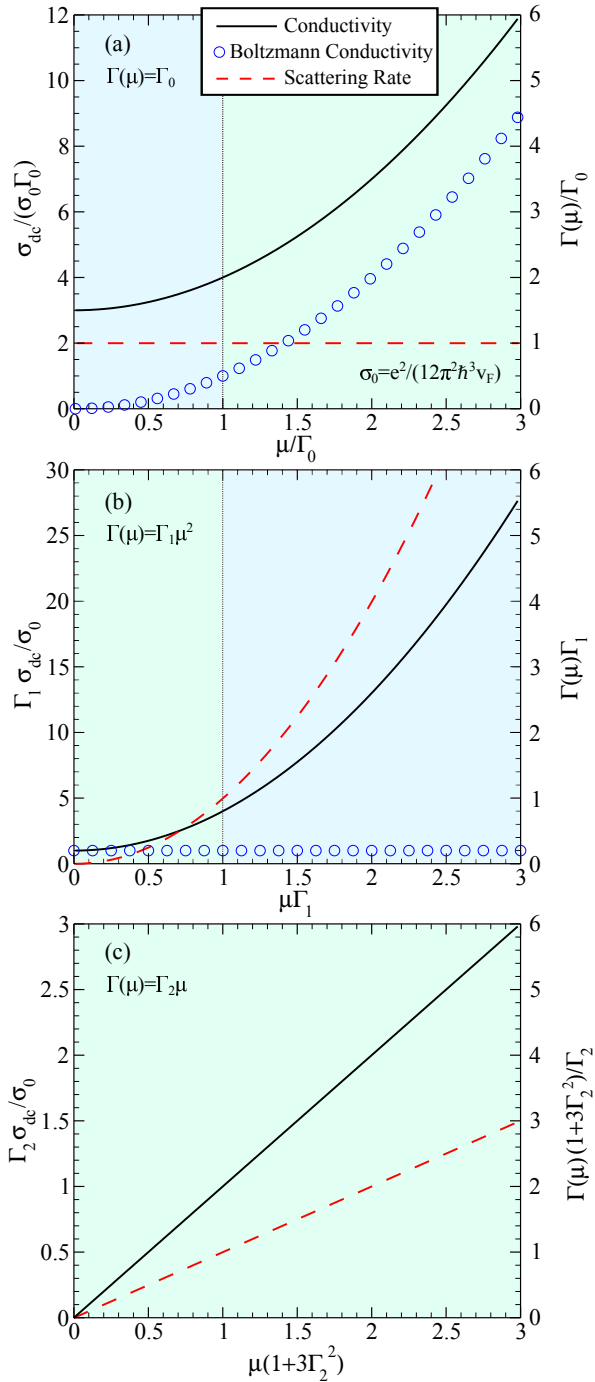}
\end{center}
\caption{\label{fig:dc-mu}(Color online) Universal dc conductivity as a function of $\mu$ for three scattering models: (a) constant scattering rate, (b) weak-scattering model, and (c) long-range Coulomb impurities.  The full results (solid black) are contrasted with those obtained through the Boltzmann approach (blue open circles).  The $\Gamma(\mu)>\mu$ region is shaded blue while $\mu>\Gamma(\mu)$ is coloured green. The scattering rate as a function of $\mu$ (dashed red) is identified on the right $y$-axis.
}
\end{figure}

For a constant scattering rate, the second term of Eqn.~\eqref{dc-T0} (proportional to $\Gamma_0$) dominates the approach to charge neutrality which corresponds to $\mu\rightarrow 0$.  This is shown in Fig.~\ref{fig:dc-mu}(a).
The solid black line gives the conductivity in units of $\sigma_0\Gamma_0$ as a function of $\mu/\Gamma_0$.  This result is to be compared with the open blue circles which give $\sigma_{\rm dc}$ as a function of $\mu$ when only the first term in Eqn.~\eqref{dc-T0} is used (Boltzmann limit).  For $\mu\rightarrow 0$, $\sigma_{\rm dc}/(\sigma_0\Gamma_0)$ takes a value of 3 while using the Boltzmann equation alone would give zero.  Note that the dc conductivity at the Dirac point is not universal as it would be in graphene but is directly proportional to $\Gamma_0$.  For finite scattering, it is always finite while the Boltzmann approach gives zero.

Weak scattering is considered in Fig.~\ref{fig:dc-mu}(b).  In contrast to $\Gamma(\omega)=\Gamma_0$, as the Dirac point is approached in the limit $\mu\rightarrow 0$, it is the Boltzmann term of Eqn.~\eqref{dc-T0} that dominates and gives the charge neutral value of the conductivity ($\sigma_0/\Gamma_1$).  Comparing the solid black curve (which contains both terms) with the open blue circles of the Boltzmann alone, we see that it is the second term of Eqn.~\eqref{dc-T0} which controls the deviation away from $\mu=0$ and gives a $\mu^2$ correction to the $\mu=0$ value.  The dashed red curve gives the $\mu$ dependence of the scattering rate in the normalized units indicated on the right vertical axis.  For the constant scattering rate model, the approach to the Dirac point falls in the region where $\Gamma(\mu)/\mu>1$ (blue shaded region) while for the weak scattering case, it occurs when $\Gamma(\mu)/\mu<1$ (green shaded region).  This is consistent with our finding that the Boltzmann approach gives the correct limit at the Dirac point in the latter model but not in the former.  The Boltzmann description is expected to apply when $\Gamma(\mu)<\mu$ and not when $\Gamma(\mu)>\mu$.   

A different pattern emerges when long-range impurities are considered [shown in Fig.~\ref{fig:dc-mu}(c)].  In that case, the condition $\Gamma(\mu)<\mu$ always applies as we expect $\Gamma_2\ll 1$.  In this model, the conductivity (solid black line) is linear in $\mu$ and the dc value at $\mu=0$ is indeed zero.  Also, the second term in Eqn.~\eqref{dc-T0} is never important; it simply changes the slope of the black line by a small amount.  

To summarize at this point, it is clear from these considerations that the Kubo formula approach to the dc conductivity can give significant corrections to the usual Boltzmann method.  Also, the different scattering models can drastically affect the approach to the Dirac point (both the value of $\sigma_{\rm dc}$ at $\mu=0$ and the power law in $\mu$ with which the asymptotic value is reached for $\mu\rightarrow 0$). For the long-range Coulomb model, the impurity scattering rate is constrained by identifying the number of scattering centres with the number of holes described by a negative value of $\mu$.  In the other models, no such constraint is imposed; therefore, from here on, we will present results mainly for the first two models which we denote model I ($\Gamma(\omega)=\Gamma_0$) and model II ($\Gamma(\omega)=\Gamma_1\omega^2$).

Returning to Eqn.~\eqref{dc}, the finite temperature conductivity is given by
 \begin{align}
\sigma_{\rm dc}=\sigma_0\int_{-\infty}^\infty d\omega\frac{1}{4T{\rm cosh}^2\left(\frac{\omega-\mu }{2T}\right)}\left[\frac{\omega^2}{\Gamma(\omega)}+3\Gamma(\omega)\right].
\end{align}
This gives [see Eqns.~\eqref{dcT-app1}, \eqref{F-app}, and ~\eqref{dcT-app2}]
\begin{align}\label{dc-T-const}
\sigma_{\rm dc}=\sigma_0\left[3\Gamma_0+\frac{1}{\Gamma_0}\left(\mu^2+\frac{1}{3}\pi^2T^2\right)\right]
\end{align}
and
\begin{align}\label{dc-T-sr}
\sigma_{\rm dc}=\sigma_0\left[3\Gamma_1\left(\mu^2+\frac{1}{3}\pi^2T^2\right)+\frac{1}{\Gamma_1}\right]
\end{align}
for models I and II, respectively.  These properly reduce to the results of Fig.~\ref{fig:dc-mu} when $T=0$.  For finite $\mu$ and $T$, the correction to the Dirac limit goes like $\mu^2$ and $T^2$, respectively.  It is of interest to compare these results with the equivalent case of graphene.  Graphene is the 2D version of a 3D Dirac Hamiltonian.  Results for the constant scattering rate $\Gamma(\omega)=\Gamma_0$
 have been given by Carbotte \emph{et al.}\cite{Carbotte:2010}. They find
 \begin{align}
 \sigma_{\rm dc}^{\rm 2D}=\frac{4e^2}{\pi h}\left[1+\frac{3}{4}\left(\frac{\mu}{\Gamma_0}\right)^2\right]
 \end{align}
for $T=0$ and finite $\mu$, and
 \begin{align}
 \sigma_{\rm dc}^{\rm 2D}=\frac{4e^2}{\pi h}\left[1+\frac{\pi^2}{9}\left(\frac{T}{\Gamma_0}\right)^2\right]
 \end{align}
for $\mu=0$ and finite $T$.  In contrast to the 3D system, the limit $T\rightarrow 0$ or $\mu\rightarrow 0$ leads to a universal minimum conductivity which is independent of the scattering rate and equal to $4e^2/(\pi h)$.  For the constant scattering rate in 3D, we get $\sigma_{\rm dc}(T\rightarrow 0)=3\sigma_0\Gamma_0$ where $\Gamma_0$ does not drop out.  However, $\sigma_{\rm dc}(T\rightarrow 0)/\Gamma_0$ is universal for a given $v_F$, independent of the scattering rate.

\begin{figure}[h!]
\begin{center}
\includegraphics[width=1.0\linewidth]{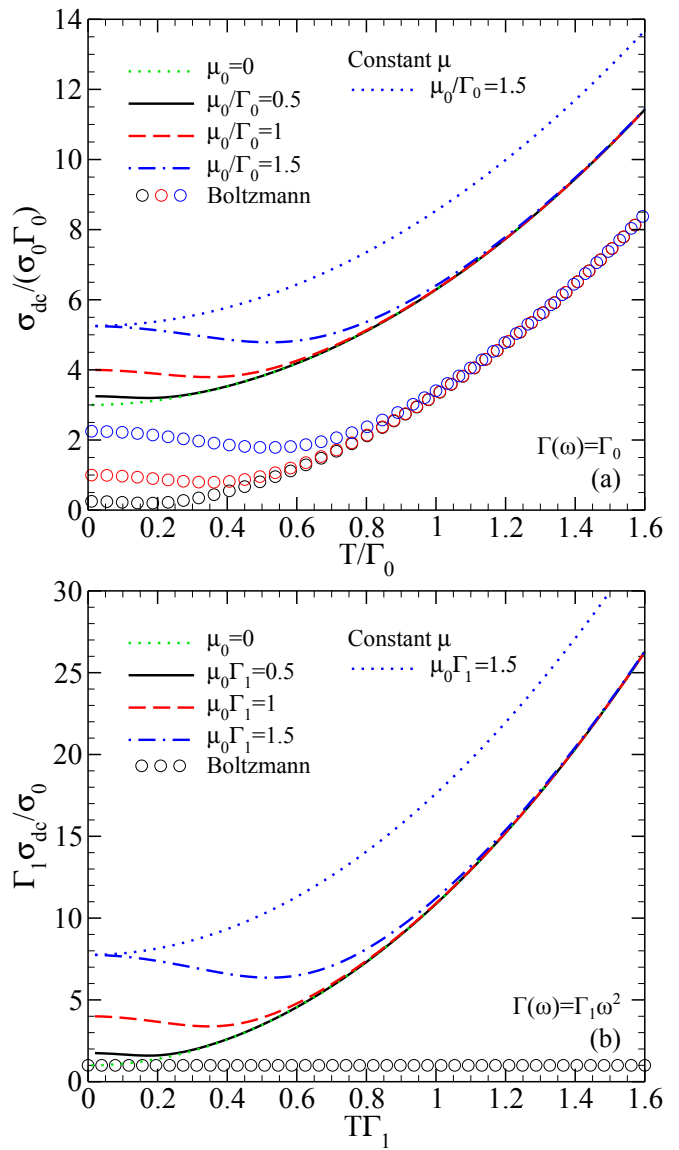}
\end{center}
\caption{\label{fig:dc-T}(Color online) Universal dc conductivity as a function of $T$ for (a) a constant scattering rate, and (b) short-range impurities.  Various values of $\mu_0$ are considered and compared to the results of the Boltzmann model (open circles) which are shifted down by three units.  The effect of the $T$ dependence of $\mu$ is seen by comparing the dash-dotted blue and dotted blue (constant $\mu$) curves.
}
\end{figure}

\begin{figure}[h!]
\begin{center}
\includegraphics[width=1.0\linewidth]{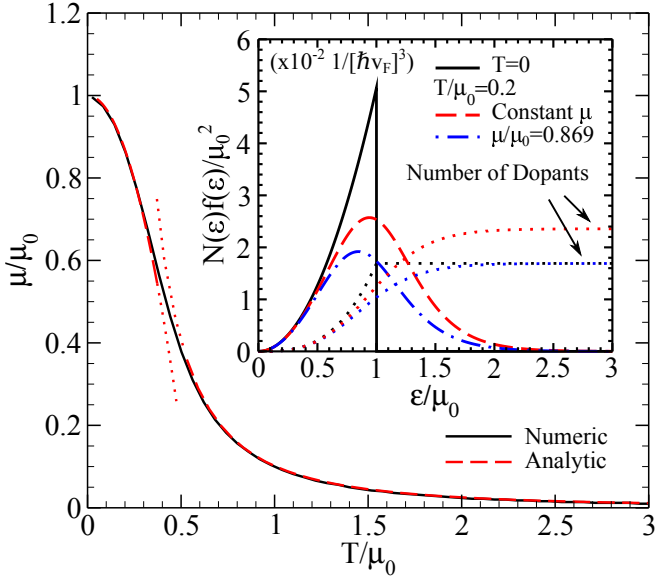}
\end{center}
\caption{\label{fig:mu}(Color online) Temperature dependence of the chemical potential.  Inset: density of states times the occupation function.  The integrated values (dotted curves) show that $\mu$ must decrease with increasing $T$ to keep the number of dopants constant.
}
\end{figure}

Results for the temperature dependence of the conductivity are presented in Fig.~\ref{fig:dc-T}.
Models I and II are shown in frames (a) and (b), respectively.  Again, scalings have been introduced on both axis so that a single curve labelled by $\mu_0/\Gamma_0$ [frame (a)] or $\mu_0\Gamma_1$ [frame (b)] applies to any value of $\Gamma_0$ or $\Gamma_1$, respectively.  In frame (a), the dotted green curve is for $\mu_0=0$.  Solid black is for $\mu_0/\Gamma_0=0.5$, dashed red corresponds to $\mu_0/\Gamma_0=1$, and dash-dotted blue is for $\mu_0/\Gamma_0=1.5$.  We also present the open-circle curves for when only the Boltzmann term in Eqn.~\eqref{dc-T-const} is used (i.e. the last term which is $\propto 1/\Gamma_0$).  The colours are set to match the same $\mu_0/\Gamma_0$ values as the full result.  The Boltzmann curves are displaced downward by a constant value of 3.  Note that the conductivity can decrease with increasing $T$, reach a minimum, and after this, start to rise as $T^2$.  The occurrence of a minimum is entirely due to the temperature dependence of the chemical potential $\mu(T)$ which reflects the energy dependence of the underlying density of states in a DSM which goes like $\omega^2$.  This is illustrated in Fig.~\ref{fig:mu} which shows how $\mu(T)$ is reduced from its $T=0$ value of $\mu_0$ as the temperature rises. 
This behaviour is determined by the conservation of particle number
\begin{align}
n=\int_0^\infty N(\omega)\left[f(\omega)-f(\omega+2\mu)\right]d\omega,
\end{align}
where
\begin{align}
N(\omega)=\frac{\omega^2}{2\pi^2\hbar^3v_F^3},
\end{align}
and can be captured by the simple analytic expression (dashed/dotted red curve in Fig.~\ref{fig:mu})
\begin{align}\label{mu}
\mu(T)=\left\lbrace\begin{array}{cc}
 \displaystyle\mu_0-\frac{\pi^2}{3}\frac{T^2}{\mu_0} &, T\ll\mu_0\\
 \displaystyle\frac{\mu_0^3}{\pi^2T^2} &, T\gg\mu_0
\end{array}\right..
\end{align}
This ensures that the number of dopant charges remains constant.  If the density of states were constant, there would be no such shift in $\mu$.  There is particle-hole symmetry and the thermal factor reduces the occupation below $\mu$ in the same proportion as it increases it above $\mu$.  To conserve particles, it is the product of the density of states times the thermal factor which enters.  This is increased more above $\mu$ than it is decreased below.  If we leave this effect out, and take $\mu(T)=\mu_0$ for the dash-dotted blue curve of Fig.~\ref{fig:dc-T}(a), we get the dotted blue curve which shows no dip as a function of $T$ but instead, increases monotonically.  The situation is similar for weak scattering shown in frame (b).  It is important to note that in this model, the Boltzmann-only solution (black open circles) is constant for all $T$ and $\mu$ and is drastically different than the full result.

\subsection{Wiedemann-Franz Law}

We now turn to a discussion of the thermal conductivity ($\kappa$), thermopower ($S$, or Seebeck coefficient) and Lorenz number [$L=\kappa/(T\sigma_{\rm dc})$].  The necessary Kubo formulas for $S$ and $\kappa$ are given in App.~\ref{app:B}.  After some straightforward algebra, we arrive at the formula for the Lorenz number.  For model I ($\Gamma(\omega)=\Gamma_0$), we find [Eqn.~\eqref{L-M1-appB}]
\begin{align}\label{L-const}
L=L_0&\left\lbrace\frac{3(15\Gamma_0^2+5\mu^2+7\pi^2T^2)}{5(9\Gamma_0^2+3\mu^2+\pi^2T^2)}\right.\notag\\
&-\left.\frac{12\mu^2\pi^2T^2}{(9\Gamma_0^2+3\mu^2+\pi^2T^2)^2}\right\rbrace,
\end{align}
where $L_0=\pi^2/(3e^2)$.  For model II ($\Gamma(\omega)=\Gamma_1\omega^2$), from Eqn.~\eqref{L-M2-appB},
\begin{align}\label{L-sr}
L=L_0&\left\lbrace\frac{5+3\Gamma_1^2(5\mu^2+7\pi^2T^2)}{5(1+\Gamma_1^2[3\mu^2+\pi^2T^2])}\right.\notag\\
&-\left.\frac{12\pi^2\Gamma_1^4\mu^2T^2}{(1+\Gamma_1^2[3\mu^2+\pi^2T^2])^2}\right\rbrace.
\end{align}
In both cases, the second term in the bracket of the expression is $S^2/L_0$
(see Appendix B). 
We have verified that when the terms proportional to $\Gamma_0$ and $\Gamma_1$
in Eqs.~\eqref{L-const} and \eqref{L-sr}, 
respectively, are dropped, these equations  reduce 
to the results found in Ref.~\cite{Lundgren:2014} which were based on a Boltzmann equation 
approach.

\begin{figure}[h!]
\begin{center}
\includegraphics[width=1.0\linewidth]{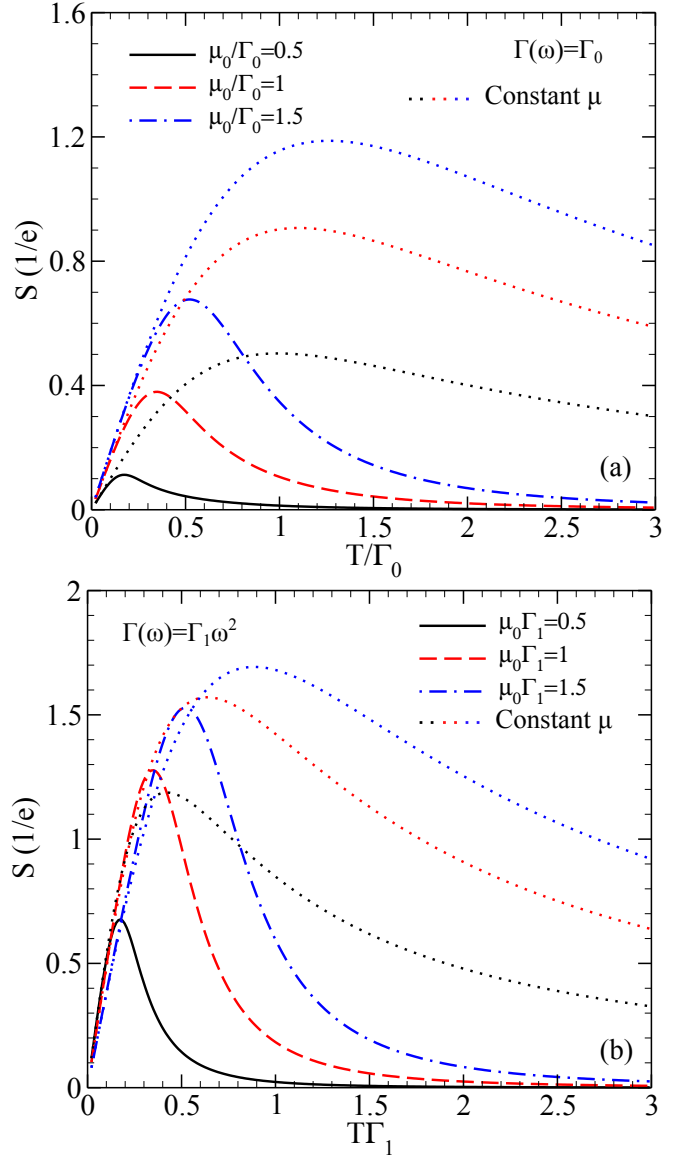}
\end{center}
\caption{\label{fig:S}(Color online) Seebeck coefficient as a function of $T$ for the (a) constant- and (b) weak-scattering models.  An increase in intensity is seen as $\mu_0$ increases.  The importance of including the temperature dependence of $\mu$ (which was done for the non-dotted curves) is emphasized by the large differences in the constant $\mu$ curves (dotted).
}
\end{figure}

Results for the thermopower are given in Fig.~\ref{fig:S}(a) and (b) for models I and II, respectively.
In both cases, we show a family of curves labelled by three values of $\mu_0/\Gamma_0$ or $\mu_0\Gamma_1$ and the temperature dependence of $\mu$ is included in the calculation.  For comparison, we also show the results when $\mu$ is kept temperature independent (dotted curves).  Clearly, the temperature dependence of $\mu$ is essential to obtaining quantitative results.  This has a strong effect on the position and sharpness of the peaks and particularly on the rapidness with which it decays for increasing $T$.

\begin{figure}[h!]
\begin{center}
\includegraphics[width=1.0\linewidth]{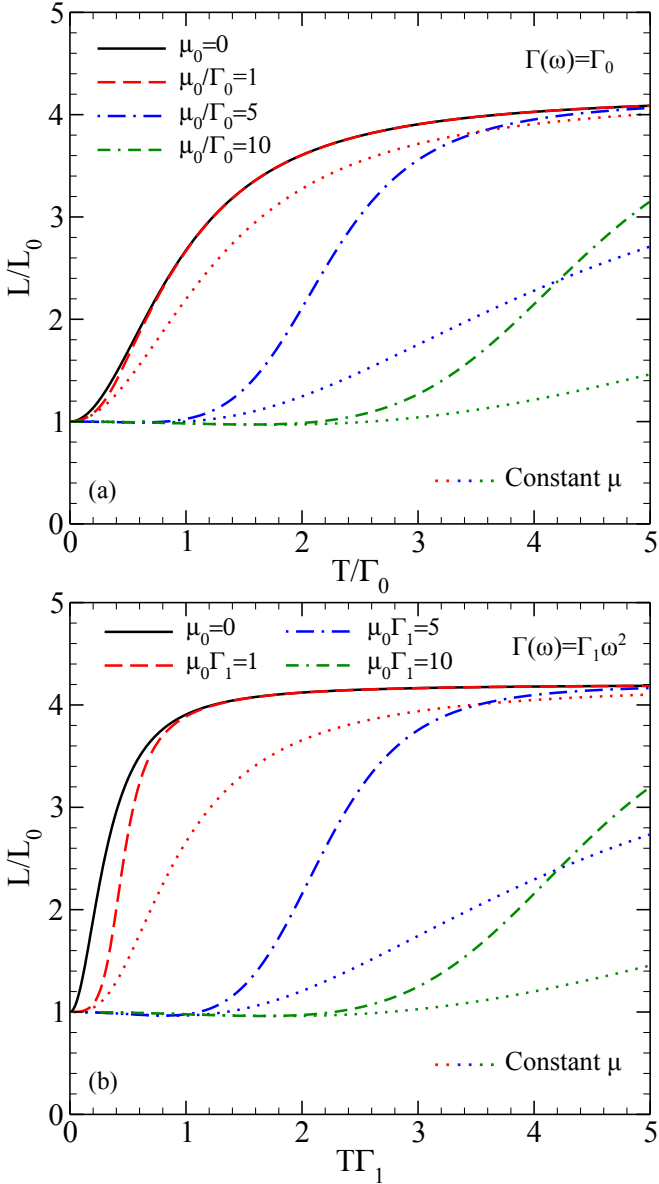}
\end{center}
\caption{\label{fig:L}(Color online) Lorenz number as a function of $T$ for the (a) constant- and (b) weak-scattering models. In all cases, $T=0$ returns the familiar value $L_0=\pi^2/(3e^2)$.  The large-$T$ limit is $L=4.2L_0$.  The rise to this limiting value shifts to higher $T$ for larger $\mu_0$.  Again, neglecting the $T$-dependence of $\mu$ has a qualitative effect on the results (dotted curves). 
}
\end{figure}

In Fig.~\ref{fig:L}, we present results for the Wiedemann-Franz law.
The Lorenz number, normalized to $L_0$ is plotted as a function of $T$ in our usual units, chosen to make the family of curves labelled by $\mu_0$ independent of $\Gamma_0$ or $\Gamma_1$.  Frame (a) displays the results for a constant scattering rate and frame (b), for weak scattering.  In all cases, regardless of the value of $\mu_0$ or the type of scattering, $L$ starts at $L_0$ when $T=0$ and rises to a value of 4.2$L_0$ as $T$ increases.  It is important to note that the temperature dependence of $\mu$ also plays an important role in determining the rise to 4.2$L_0$.  This is seen in Fig.~\ref{fig:L} by comparing the full result to the case when $\mu(T)=\mu_0$ (dotted curves).  Differences can be large.

The results in Fig.~\ref{fig:L} can be understood by examining Eqns.~\eqref{L-const} and \eqref{L-sr}.  Firstly, at $T=0$, the thermopower correction to the Wiedemann-Franz law (second term) vanishes and we get
\begin{align}\label{L-const-T0}
L&=L_0\left\lbrace\frac{3\Gamma_0^2+\mu^2}{3\Gamma_0^2+\mu^2}\right\rbrace=L_0,
\end{align}
for $\Gamma(\omega)=\Gamma_0$, and
\begin{align}\label{L-sr-T0}
L&=L_0\left\lbrace\frac{3\Gamma_1^2\mu^2+1}{3\Gamma_1^2\mu^2+1}\right\rbrace=L_0,
\end{align}
for $\Gamma(\omega)=\Gamma_1\omega^2$. These reduce to $L_0$ regardless of the scattering parameter or chemical potential.  Further, when $T$ becomes the dominant scale, the second terms of Eqns.~\eqref{L-const} and \eqref{L-sr} can be dropped as they go like $1/T^2$ and we are left with
\begin{align}
L=\frac{21}{5}L_0=4.2L_0
\end{align}
in both models.

\begin{figure}[h!]
\begin{center}
\includegraphics[width=1.0\linewidth]{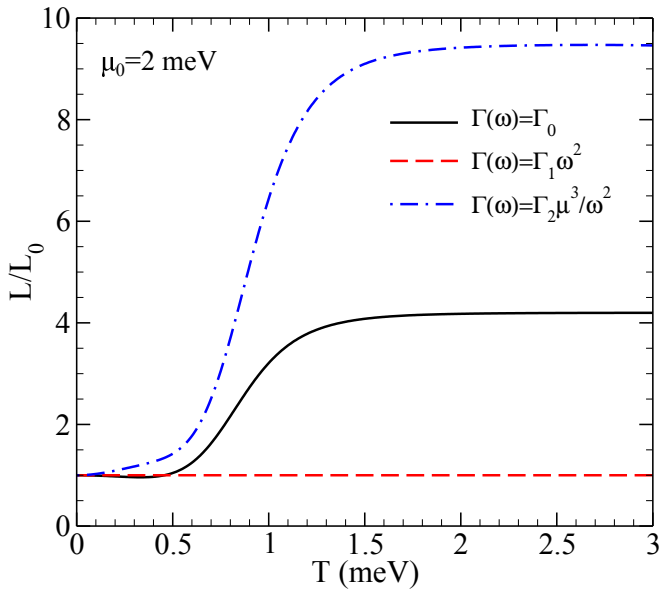}
\end{center}
\caption{\label{fig:L-Boltz}(Color online) Lorenz number as a function of $T$ calculated from the Boltzmann equation.  Three models are considered: constant-scattering, weak-scattering, and long-range Coulomb impurities. As expected, the $T=0$ limit is $L_0=\pi^2/(3e^2)$.  The large-$T$ limit is different for all three models; it is $L=4.2L_0$, $L_0$ and $9.49L_0$, respectively.
}
\end{figure}

It is also interesting to note that the numerators of the second terms of Eqns.~\eqref{L-const} and \eqref{L-sr} have different origins.  For model I [Eqn.~\eqref{L-const}], it comes entirely from the Boltzmann contribution.  For model II [Eqn.~\eqref{L-sr}], it arises due to the correction piece proportional to $\Gamma_1$ rather than the $1/\Gamma_1$ term of Boltzmann theory. The first term in the Lorenz ratio, of course, has contributions from both as does the dc conductivity denominator of the second terms in Eqns.~\eqref{L-const} and \eqref{L-sr}.  If we retain only the Boltzmann contribution, we obtain Fig.~\ref{fig:L-Boltz}.  
Now, Eqn.~\eqref{L-sr} gives $L=L_0$ for all $T$ (dashed red).  It is clearly the non-Boltzmann piece that provides the temperature dependence of $L$ for weak scattering.  For constant impurity scattering, the rise to $4.2L_0$ is seen even in the Boltzmann approach (solid black).  Returning to the long-range Coulomb impurity model, we find that the pure Boltzmann approach gives
\begin{align}\label{L-Born}
L&=L_0\left\lbrace\frac{1+\frac{42}{5}\pi^2(T/\mu)^2+\frac{31}{7}\pi^4(T/\mu)^4}{1+2\pi^2(T/\mu)^2+\frac{7}{15}\pi^4(T/\mu)^4}-\frac{3}{\pi^2}\left(\frac{4\mu}{T}\mathcal{B}\right)^2\right\rbrace
\end{align}
where
\begin{align}
\mathcal{B}=\frac{\frac{1}{3}\pi^2(T/\mu)^2+\frac{7}{15}\pi^4(T/\mu)^4}{1+2\pi^2(T/\mu)^2+\frac{7}{15}\pi^4(T/\mu)^4}.
\end{align}
The $T$-dominated limit of Eqn.~\eqref{L-Born} is particularly interesting.  It gives
\begin{align}
L=9.49L_0.
\end{align}
This yields a different plateau in the Lorenz ratio.  Numerical results are shown in Fig.~\ref{fig:L-Boltz} (dash-dotted blue).  Clearly, in the Boltzmann approach, the three scattering models give three different saturated values at high $T$.

\section{AC Conductivity}

The general formula for the absorptive part of the dynamical conductivity is given in App.~\ref{app:A}.  In particular, the intra- and interband contributions for any scattering model are given by Eqns.~\eqref{intra-app} and \eqref{inter-app}, respectively.  For a constant $\Gamma(\omega)=\Gamma_0$, we find
\begin{align}\label{inter-const}
&\sigma_{xx}^{\rm IB}(\Omega)=\frac{e^2}{3\pi^2\hbar^3 v_F}\int_{-\infty}^\infty d\omega\frac{f(\omega)-f(\omega+\Omega)}{\Omega}\notag\\
&\times\frac{\Gamma_0\left(\omega^2+(\omega+\Omega)^2+2\Gamma_0^2\right)}{4\Gamma_0^2+(2\omega+\Omega)^2},
\end{align}
and
\begin{align}\label{intra-const}
&\sigma_{xx}^{\rm D}(\Omega)=\frac{e^2}{6\pi^2\hbar^3 v_F}\int_{-\infty}^\infty d\omega\frac{f(\omega)-f(\omega+\Omega)}{\Omega}\notag\\
&\times\frac{\Gamma_0\left(\omega^2+(\omega+\Omega)^2+2\Gamma_0^2\right)}{4\Gamma_0^2+\Omega^2},
\end{align}
for the inter- and intraband components, respectively.  In the clean limit ($\Gamma_0\rightarrow 0$), these simplify greatly and analytic results can be obtained:
\begin{align}\label{inter-G0}
\sigma_{xx}^{\rm IB}(\Omega)&=\frac{e^2}{24\pi\hbar^3 v_F}\left[\frac{f(-\frac{\Omega}{2})-f(\frac{\Omega}{2})}{\Omega}\right]\Omega^2\notag\\
&=\frac{e^2}{24\pi\hbar^3 v_F}\Omega\frac{{\rm sinh}(\beta\Omega/2)}{{\rm cosh}(\beta\mu)+{\rm cosh}(\beta\Omega/2)},
\end{align}
where $\beta=1/T$, and
\begin{align}\label{intra-G0}
\sigma_{xx}^{\rm D}(\Omega)&=\frac{e^2}{6\pi\hbar^3 v_F}\delta(\Omega)\int_{-\infty}^\infty \left(-\frac{\partial f(\omega)}{\partial\omega}\right)\omega^2\notag\\
&=\frac{e^2}{6\pi\hbar^3 v_F}\delta(\Omega)\left(\mu(T)^2+\frac{\pi^2}{3}T^2\right).
\end{align}
It is important to connect these results for $\sigma_{xx}^{\rm IB}(\Omega)$ and $\sigma_{xx}^{\rm D}(\Omega)$ obtained in the clean limit with our previous results for the dc conductivity of Eqn.~\eqref{dc-T-const}.  In deriving the interband and intraband contributions to the dynamical conductivity [Eqns.~\eqref{inter-G0} and ~\eqref{intra-G0}, respectively] we have kept only the leading term as $\Gamma_0\rightarrow 0$ which goes like $1/\Gamma_0$.  Effectively, we have dropped the first term in Eqn.~\eqref{dc-T-const} which is $3\Gamma_0\sigma_0$ and is only important if both $\mu$ and $T$ are zero.  Therefore, only the Boltzmann terms are retained.  Clearly, as $\Omega\rightarrow 0$, the interband contribution $\sigma_{xx}^{\rm IB}(\Omega\rightarrow 0)=0$ and the intraband component $\sigma_{xx}^{\rm D}(\Omega\rightarrow 0)=(\sigma_0/\Gamma_0)(\mu^2+\pi^2T^2/3)$ which agrees with Eqn.~\eqref{dc-T-const}.  Here, we have replaced the Dirac $\delta$-function in Eqn.~\eqref{intra-G0} by a Lorentzian of width $2\Gamma_0$ and taken the dc limit $\Omega\rightarrow 0$ before considering the clean limit $\Gamma_0\rightarrow 0$ and retaining only the leading term $\propto 1/\Gamma_0$.

The optical spectral weight under the Drude is defined as
\begin{align}\label{spec-Drude}
\int_{0^+}^\infty\sigma_{xx}^{\rm D}(\Omega)d\Omega &\equiv W_{\rm D}(T,\mu)\notag\\
&=\frac{e^2}{12\pi\hbar^3 v_F}\left(\mu(T)^2+\frac{\pi^2}{3}T^2\right)
\end{align}
and depends on both the chemical potential and temperature.  In general, the chemical potential has a $T$ dependence.  At low temperature, it is given by the approximate expression [see Eqn.~\eqref{mu}] 
\begin{align}
\mu(T)=\mu_0\left[1-\frac{\pi^2}{3}\left(\frac{T}{\mu_0}\right)^2\right],
\end{align}
where $\mu_0$ is its zero-temperature value.  Therefore, in this limit,
\begin{align}
W_{\rm D}(T,\mu)=\frac{e^2}{12\pi\hbar^3 v_F}\left(\mu_0^2-\frac{\pi^2}{3}T^2+\frac{\pi^4}{9}\frac{T^4}{\mu_0^2}\right),
\end{align}
so that $W_{\rm D}(T,\mu)$ will first decrease as $T$ increases.  For large $T$, 
\begin{align}
\mu(T)=\frac{\mu_0^3}{\pi^2T^2},
\end{align}
and
\begin{align}
W_{\rm D}(T,\mu)=\frac{e^2}{12\pi\hbar^3 v_F}\left(\frac{\mu_0^6}{\pi^4T^4}+\frac{\pi^2}{3}T^2\right).
\end{align}
Therefore, in the large $T$ limit, $W_{\rm D}(T,\mu)$ will increase as $T^2$.  The corresponding effect has been noted in experiments on graphene\cite{Frenzel:2014} where an optical pumping terahertz technique was used and the optical spectral weight of the Drude was obtained over a large $T$ interval.  They saw the initial decrease of spectral weight as $T$ is increased out of $T=0$.  The expected trend towards linearity in $T$ for 2D, which is predicted by theory\cite{Gusynin:2009}, was observed.  There is also preliminary data that bares on this issue in DSMs.  Examples include the optical work by Sushkov \emph{et al.}\cite{Sushkov:2015} in the pyrochlore iridates Eu$_2$Ir$_2$O$_7$, Chen \emph{et al.}\cite{Chen:2015} in ZrTe$_5$ and Xu \emph{et al.}\cite{Xu:2015} in TaAs.

\begin{figure}[h!]
\begin{center}
\includegraphics[width=1.0\linewidth]{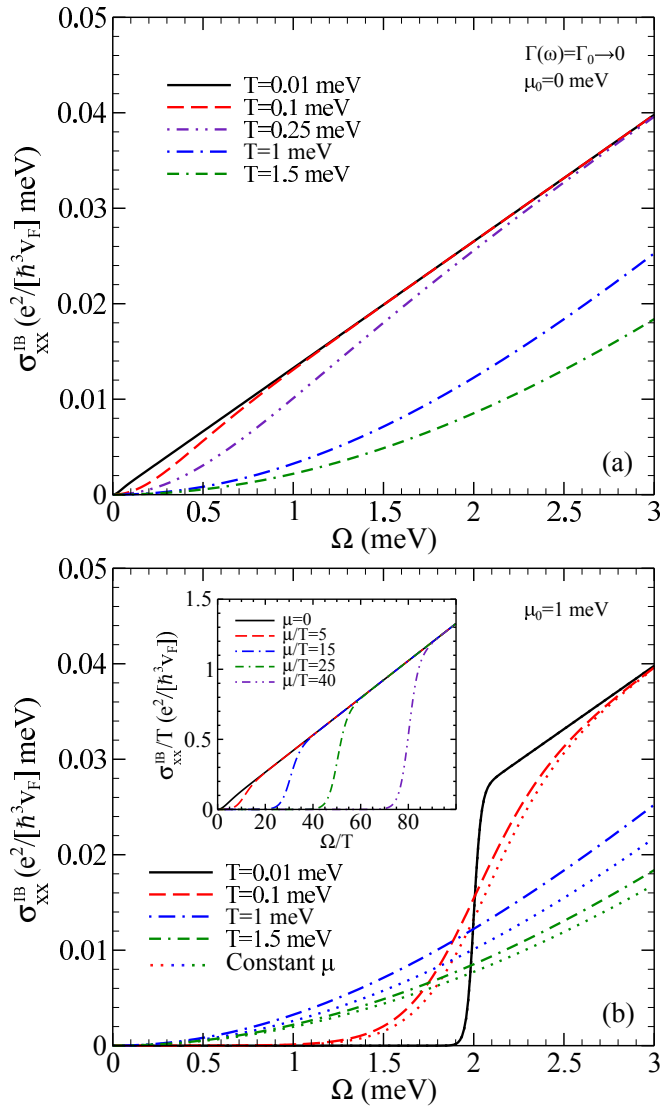}
\end{center}
\caption{\label{fig:IB}(Color online) Interband conductivity for $\Gamma(\omega)=\Gamma_0\rightarrow 0$.  (a) For $\mu_0=0$, varying $T$ yields a reduction in the low-frequency response before it returns to the $T=0$ value at large $\Omega$. (b) When $\mu_0$ is finite, a step is seen in $\sigma_{xx}^{\rm IB}$ at $\Omega=2\mu$, after which, the $\mu_0=0$ results are retained.  This step becomes increasingly washed-out as $T$ increases.  Inset: $\sigma_{xx}^{\rm IB}/T$ for large values of $\mu/T$.  The step at $\Omega=2\mu$ becomes sharper as $\mu$ increases relative to $T$.  A linear high-$\Omega$ response is clear.
}
\end{figure}

In Fig.~\ref{fig:IB}, we show results for the interband contribution to the conductivity given by Eqn.~\eqref{inter-G0} at various temperatures.  
Frame (a) displays $\mu_0=0$, while frame (b) contains results for $\mu_0=1$ meV.  In frame (a), solid black corresponds to $T=0.01$ meV, dashed red to $T=0.1$ meV, dash-double-dotted purple to $T=0.25$ meV, dash-dotted blue to $T=1$ meV and double-dash-dotted green to $T=1.5$ meV.  At low temperature (black curve), we retain the expected linear behaviour since tanh$(\beta\Omega/2)\rightarrow 1$ for $\beta\rightarrow\infty$ in Eqn.~\eqref{inter-G0}.  As $T$ increases, the low $\Omega$ region of the curve is rapidly depleted due to the transfer of spectral weight from the interband to the intraband.  This is a manifestation of the reduction of interband transitions due to Pauli blocking in the conduction band and the reduction in the probability of occupation in the valence band.  In addition, the linear behaviour is only recovered at values of $\Omega$ an order of magnitude greater than $T$ (see the dash-double-dotted purple curve).  For the two higher $T$ values shown, linearity is not observed as it occurs for $\Omega$ much greater than shown here.  A similar scenario is seen in frame (b) where $\mu_0=1$ meV.  Again, several $T$ values are shown: 0.01, 0.1, 1, and 1.5 meV  which are coloured the same as in the upper frame.  For low $T$, the solid black curve displays a clear cutoff at $\Omega=2\mu$ as expected from Eqn.~\eqref{inter-G0} where the quotient of hyperbolic trigonometric functions reduces to the Heaviside step function $\Theta(\Omega-2\mu)$.  This sharp rise is quickly smeared as $T$ increases.  Even for $T=1$ meV (dash-dotted blue), the cutoff is indiscernible.  The inset of Fig.~\ref{fig:IB}(b) shows the clean limit as a function of $\Omega/T$ for various values of $\mu/T$.  Clearly, the sharpness of the step-up increases as $\mu/T$ becomes larger.  Finally, we note that the dotted lines in frame (b) correspond to $\mu$ being held temperature independent.  The colours are set to match those of the corresponding $T$ values. This does not change the main qualitative features.  

\begin{figure}[h!]
\begin{center}
\includegraphics[width=1.0\linewidth]{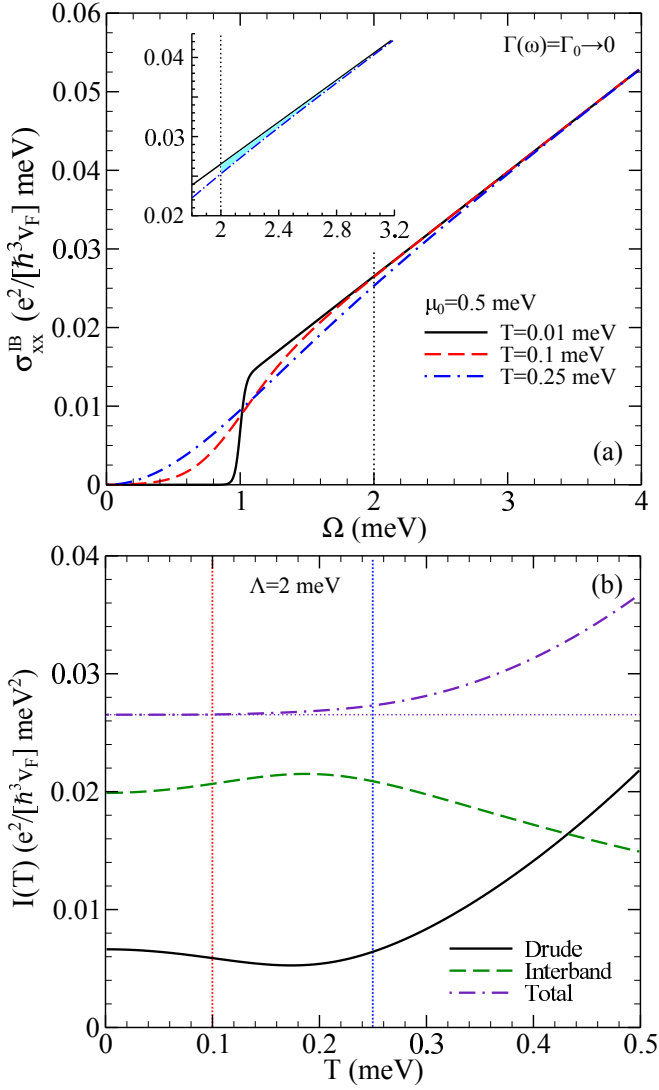}
\end{center}
\caption{\label{fig:spec}(Color online) (a) $\sigma_{xx}^{\rm IB}$ for $\Gamma(\omega)=\Gamma_0\rightarrow 0$.  Above $2\mu$, as $T$ increases, the low-frequency response is reduced from its $T=0$ value. This yields a loss in spectral weight for a low cutoff $\Lambda$ (inset). (b) Drude, interband and total optical spectral weight for $\Lambda=2$ meV and the same parameters as (a).  For low $T$, conservation of spectral weight is observed for sufficiently large $\Lambda$.
}
\end{figure}

It is clear from our discussion that the optical spectral weight under the Drude and in the interband transitions are both changed with variations in temperature.  For graphene, it was demonstrated by Gusynin \emph{et al.}\cite{Gusynin:2009} that the total spectral weight up to an energy $\Lambda\gg T$ is conserved.  We find that this is also the case for a 3D DSM as demonstrated in Fig.~\ref{fig:spec}.  
In frame (a), the interband conductivity as a function of photon frequency is shown for $\mu_0=0.5$ meV and $T=0.01$ meV (solid black), $T=0.1$ meV (dashed red), and $T=0.25$ meV (dash-dotted blue).  Frame (b) presents the optical spectral weight $I(T)$ up to $\Lambda=2$ meV where
\begin{align}\label{spec-def}
I(T)=\int_{0^+}^\Lambda\sigma_{xx}(\Omega,T)d\Omega.
\end{align}
The solid black curve gives the Drude contribution while the dashed green and dash-dotted purple curves give the interband and total weights, respectively.  At low temperature, the total spectral weight up to $\Lambda=2$ meV is completely $T$ independent.  Deviations from a constant set in at $T\approx 0.15$ meV.  This does not imply that the conservation law on optical spectral weight holds only at small $T$.  It means that conservation for larger $T$ requires a higher $\Lambda$.  This is graphically seen in Fig.~\ref{fig:spec}(a).  Consider the dotted vertical line at $\Omega=2$ meV [equal to the $\Lambda$ cutoff used in frame (b)].  For the dash-dotted blue curve, which gives the interband conductivity at $T=0.25$ meV [dotted blue vertical line in frame (b)], the cutoff at 2 meV does not fall above the point at which the conductivity has returns to its $T=0$ value.  More optical spectral weight is lost above this energy as illustrated in the inset by the shaded region.  Thus, to see the sum rule obeyed at this temperature would require a large $\Lambda$.  At $T=0.1$ meV, this cutoff is clearly adequate and indeed the sum rule is seen in the lower frame (dotted red vertical line).  In conclusion, if we take a sufficiently large cutoff (more than an order of magnitude) compared with the temperature of interest, we obtain conservation of optical spectral weight as also found in graphene.

\begin{figure}[h!]
\begin{center}
\includegraphics[width=1.0\linewidth]{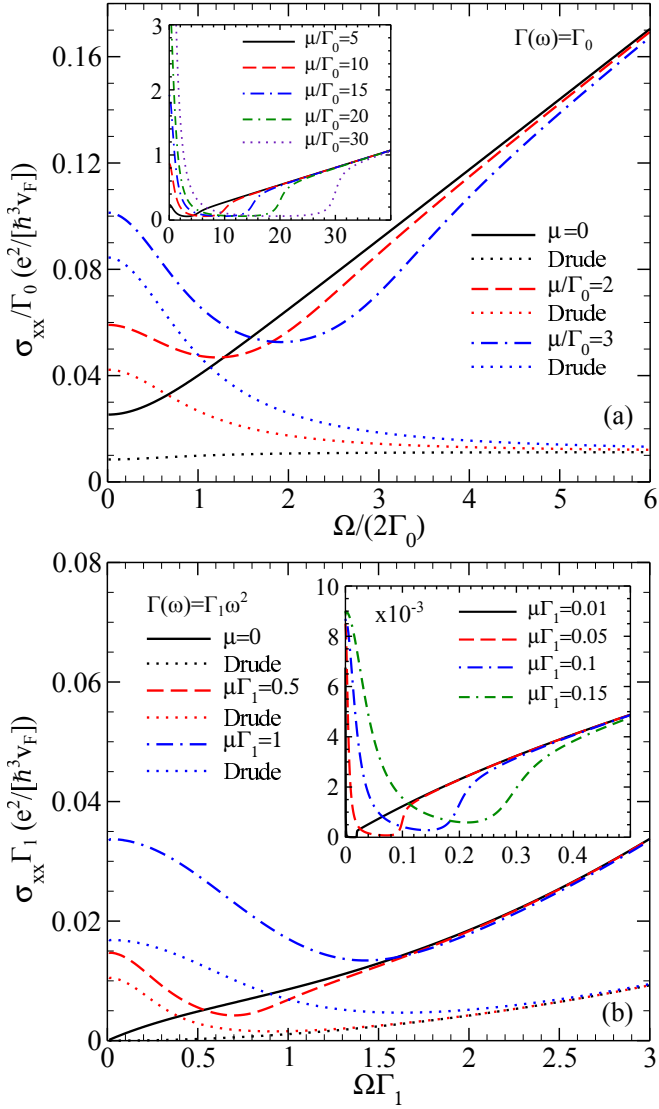}
\end{center}
\caption{\label{fig:cond-univ}(Color online) Optical conductivity for (a) $\Gamma(\omega)=\Gamma_0$ and (b) $\Gamma(\omega)=\Gamma_1\omega^2$.  (a) As $\mu$ increases, a larger Drude response is seen centred about $\Omega=0$ (dotted curves).  At higher frequency, a linear interband background emerges.  Inset: large-$\mu$ limit.  A stronger Drude is observed, before a depletion of conductivity prior to a step at $\Omega=2\mu$ into a linear background. (b)  In contrast to (a), the interband background and high-$\Omega$ Drude is quadratic due to the energy dependence of $\Gamma(\omega)$.
}
\end{figure}

Next, we consider $T=0$ but a finite scattering rate $\Gamma(\omega)$.  Only the case of $\Gamma(\omega)=\Gamma_0$ is sufficiently simple to provide a useful analytic result.  In this instance, the inter- and intraband conductivities are given by
\begin{align}
&\sigma_{xx}^{\rm IB}(\Omega)=\frac{e^2}{3\pi^2\hbar^3 v_F}\int_{\mu-\Omega}^\mu d\omega\frac{\Gamma_0}{\Omega}\frac{\omega^2+(\omega+\Omega)^2+2\Gamma_0^2}{4\Gamma_0^2+(2\omega+\Omega)^2},
\end{align}
and
\begin{align}
&\sigma_{xx}^{\rm D}(\Omega)=\frac{e^2}{6\pi^2\hbar^3 v_F}\int_{\mu-\Omega}^\mu d\omega\frac{\Gamma_0}{\Omega}\frac{\omega^2+(\omega+\Omega)^2+2\Gamma_0^2}{4\Gamma_0^2+\Omega^2},
\end{align}
respectively.  These give
\begin{align}\label{inter-G}
\sigma_{xx}^{\rm IB}(\Omega)=\frac{e^2}{6\pi^2\hbar^3 v_F}\Gamma_0&\left[1+\frac{\Omega}{4\Gamma_0}\left\lbrace{\rm arctan}\left(\frac{2\mu+\Omega}{2\Gamma_0}\right)\right.\right.\notag\\
&-\left.\left.{\rm arctan}\left(\frac{2\mu-\Omega}{2\Gamma_0}\right)\right\rbrace\right],
\end{align}
and
\begin{align}\label{intra-G}
\sigma_{xx}^{\rm D}(\Omega)=&\frac{e^2}{12\pi^2\hbar^3 v_F}\frac{\Gamma_0}{1+[\Omega/(2\Gamma_0)]^2}\notag\\
&\times\left[1+\left(\frac{\mu}{\Gamma_0}\right)^2+\frac{4}{3}\left(\frac{\Omega}{2\Gamma_0}\right)^2\right].
\end{align}
Adding Eqns.~\eqref{inter-G} and \eqref{intra-G} in the limit $\Omega\rightarrow 0$, returns the total dc conductivity of Eqn.~\eqref{dc-T-const} when $T=0$. Results applicable to the approach to the Dirac point (small $\mu$) are shown in Fig.~\ref{fig:cond-univ}(a) while large $\mu$ is considered in the inset.
In the main frame, the solid black curve gives $\mu=0$, the dashed-red shows $\mu/\Gamma_0=2$ and dash-dotted blue displays $\mu/\Gamma_0=3$.  In all three cases, the associated colour-coordinated dotted curves give the Drude contribution.  These are strikingly different from the ordinary Drude result of metal theory which are described by the simple Lorentzian form $2\Gamma_0/[(2\Gamma_0)^2+\Omega^2]$, where $2\Gamma_0$ is the transport scattering rate.  This difference is expected from Eqn.~\eqref{intra-G} where the square brackets contain additional terms proportional to $\mu^2$ and $\Omega^2$ as well as the constant which multiples the Lorentzian.  These modifications imply that the Drude conductivity does not decay like $\Omega^{-2}$ at large $\Omega$ but rather saturates at a constant value of 
\begin{align}
\sigma_D^\infty=\frac{e^2}{9\pi^2\hbar^3v_F}\Gamma_0.
\end{align}
For large $\mu/\Gamma_0$, the inset shows an increase in spectral weight under the Drude which becomes better defined about $\Omega=0$.  This is followed by a region of low conductivity before a rapid rise occurs at $\Omega=2\mu$ associated with interband transitions.  This step sharpens with increasing $\mu/\Gamma_0$, as is clearly seen in the inset for Fig.~\ref{fig:cond-univ}(a) which has results for $\mu/\Gamma_0=5$ (solid black) to $\mu/\Gamma_0=30$ (dotted purple) in increments of 5.

It is interesting to closely examine the dc limit and relate this value to our results given in Eqn.~\eqref{dc-T-const}. Referring back to Eqns.~\eqref{dc-intra-app}-\eqref{dc-app}, it can be seen that, in Eqn.~\eqref{dc-T-const}, the second term ($\propto 1/\Gamma_0$), is a Boltzmann-type term that appears only in the Drude conductivity and is zero at $T=0$ and $\mu=0$.  The first term of Eqn.~\eqref{dc-T-const} is the correction provided by the Kubo formula and is $\propto (\Gamma_0+2\Gamma_0)$ where the first $\Gamma_0$ comes from the Drude piece and $2\Gamma_0$ comes from the interband piece.  Consequently, upon examining the solid black curve of Fig.~\ref{fig:cond-univ}(a) in the limit $\Omega\rightarrow 0$, one sees that it is three times the corresponding Drude value (dotted black).  When $\mu\neq 0$, the Boltzmann term makes an additional contribution of $1/(12\pi^2)(\mu/\Gamma_0)^2$ in the units used here.

For the weak-scattering model where $\Gamma(\omega)=\Gamma_1\omega^2$, similar results are presented in Fig.~\ref{fig:cond-univ}(b).  These were obtained numerically from the more complicated equations which take full account of the energy dependence of the impurity scattering [Eqns.~\eqref{intra-app} and \eqref{inter-app}].  The format is the same as Fig.~\ref{fig:cond-univ}(a).  In the main frame, we show results for the real part of the dynamical conductivity which are normalized to be independent of $\Gamma_1$.  Three values of $\mu\Gamma_1$ are plotted: 0 (solid black), 0.5 (dashed red) and 1 (dash-dotted blue).  Again, the Drude contribution is given by the colour-coordinated dotted lines.  For $\mu=0$, no Drude peak is seen at $\Omega=0$ but there is still an intraband contribution (dotted black) which starts from zero at $\Omega=0$ and rises with increasing $\Omega$. This behaviour is traced to the vanishing of the scattering rate at $\omega=0$ and its quadratic growth in $\omega$ ($\Gamma(\omega)=\Gamma_1\omega^2$).  For finite $\mu$, there is a Drude peak centred at $\Omega=0$ but the intraband conductivity eventually flattens out before it begins to increase in a quasilinear fashion.  The total conductivity shows a Drude-like peak as $\Omega\rightarrow 0$ which is followed by a region which is nearly linear in $\Omega$.  

It is important to note that in frame (a), $\mu/\Gamma_0$ is used while the scaling in (b) is $\mu\Gamma_1$.  In terms of the scattering rate at the chemical potential,  $\mu/\Gamma_0=\mu/\Gamma(\mu)$ while $\mu\Gamma_1=\Gamma(\mu)/\mu$.  Thus, in the upper frame, the Boltzmann regime of $\mu\gg\Gamma(\mu)$ corresponds to large values of $\mu/\Gamma_0$ while in the lower frame, it is associated with small values of $\mu\Gamma_1$.  This is shown in the inset of Fig.~\ref{fig:cond-univ}(b) for four values of $\mu\Gamma_1$: 0.001 (solid black), 0.005 (dashed red), 0.1 (dash-dotted blue), and 0.15 (double-dash-dotted green).  The $x$-axis is $\Omega\Gamma_1$; thus, the expected jump at $\Omega=2\mu$ corresponds to $2\mu\Gamma_1$ on this axis.  Therefore, the solid black curve jumps at 0.02, while the double-dash-dotted green curve steps up at 0.3.  The jump is sharper for $\mu\Gamma_1=0.001$ than 0.15 as the scattering rate has increased from $\Gamma(\mu)/\mu=0.01$ to 0.15.

\begin{figure}[h!]
\begin{center}
\includegraphics[width=1.0\linewidth]{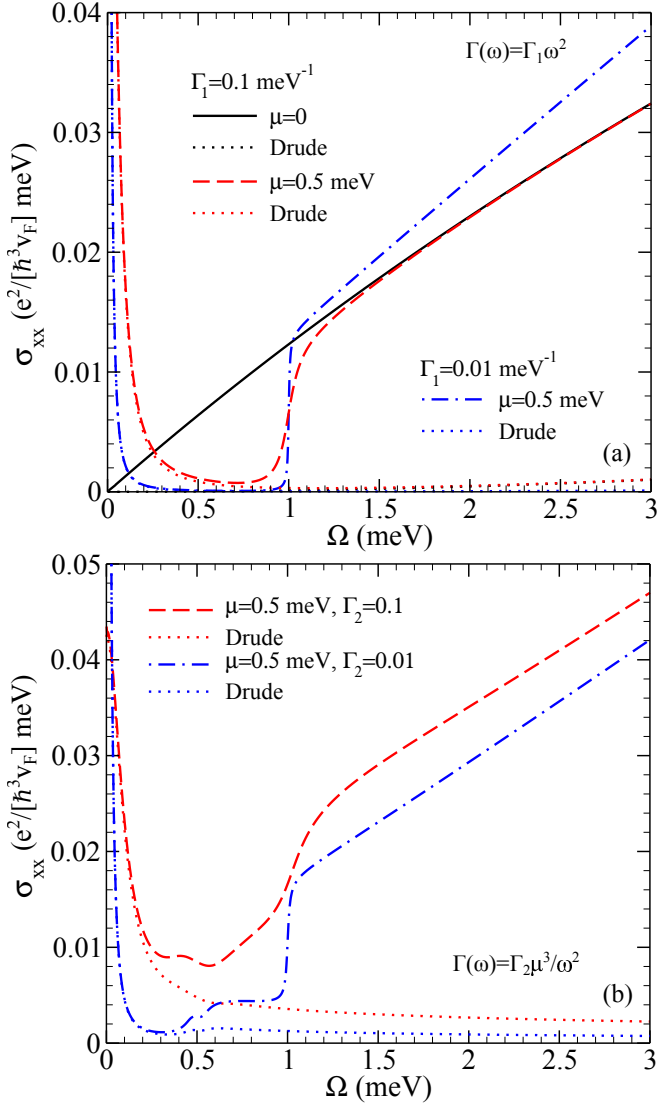}
\end{center}
\caption{\label{fig:sr-Born}(Color online) Optical conductivity for (a) $\Gamma(\omega)=\Gamma_1\omega^2$ and (b) $\Gamma(\omega)=\Gamma_2\mu^3/\omega^2$.  In both cases, the slope and height of the interband background is highly dependent on the scattering rate amplitude. 
}
\end{figure}

Figure~\ref{fig:sr-Born} emphasizes another important point about the behaviour of the conductivity when the Drude and interband regions are clearly separated with a region of low conductivity between them.  
Frame (a) shows the short-range impurity scattering model, while long-range Coulomb scattering is used in (b).  In both frames, $\mu$ is set at 0.5 meV and we, thus, expect a jump at $\Omega=1$ meV except for the solid black curve in frame (a) which has no charge doping ($\mu=0$) and is for comparison.  We take $\Gamma_1=0.1$ meV$^{-1}$ [frame (a)] and $\Gamma_2=0.1$ [frame (b)] for the solid black and dashed red curves.  The dash-dotted blue curve corresponds to $\Gamma_1=0.01$ meV$^{-1}$ [frame (a)] and $\Gamma_2=0.1$ [frame (b)].  In both models, the jump at $\Omega=2\mu=1$ meV is sharpest in the dash-dotted blue curve which is purer.  More prominent structures are seen in the gap of the long-range Coulomb conductivity as compared to model II.  This is expected since the optical conductivity involves a product of the density of states factor $\omega^2$ and the scattering rate $\Gamma(\omega+\Omega)$ which is very different in these two cases.  More precisely, the factor $\Gamma(\omega)(\omega+\Omega)^2+\Gamma(\omega+\Omega)\omega^2$, is equal to $\Gamma_1(\omega^2(\omega+\Omega)^2+(\omega+\Omega)^2\omega^2)$ for $\Gamma(\omega)=\Gamma_1\omega^2$.  For $\Gamma(\omega)=\Gamma_2\mu^3/\omega^2$, it is the more complicated $\Gamma_2\mu^3[(\omega+\Omega)^2/\omega^2+\omega^2/(\omega+\Omega)^2]$ which introduces significant structure in the Fig.~\ref{fig:sr-Born}(b).  We see in the colour-coordinated dotted curves for the Drude contribution alone that impurity structures are present in the intraband absorption processes as well.  These are traced to the energy dependence of the scattering rate.  Finally, in the region of photon energy shown, when the scattering rate increases with $\omega$ [$\Gamma(\omega)=\Gamma_1\omega^2$], the cleaner sample has a conductivity that is above the dirtier system in the quasilinear region while it is the opposite when the scattering rate decreases with increasing $\omega$ ($\Gamma(\omega)=\Gamma_2\mu^3/\omega^2$).  

\section{Weyl Semimetals}

\begin{figure}[h!]
\begin{center}
\includegraphics[width=1.0\linewidth]{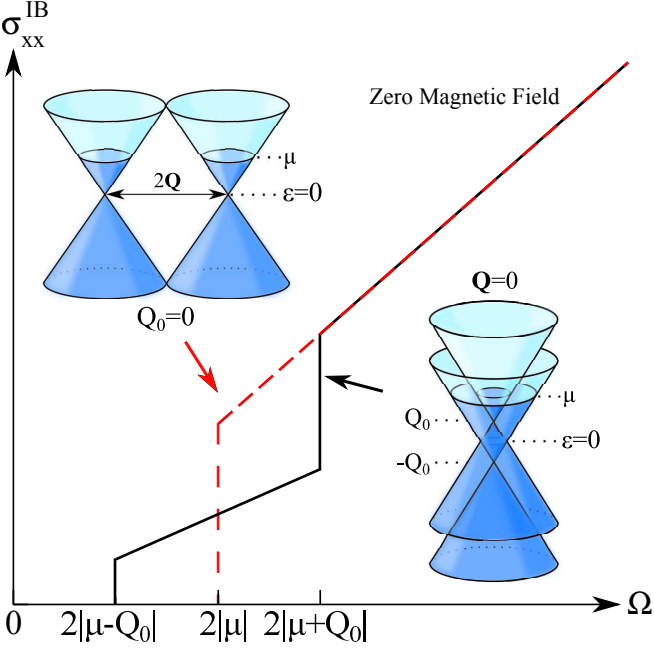}
\end{center}
\caption{\label{fig:DW}(Color online) Schematic plot of the interband conductivity for a time-reversal symmetry breaking WSM (dashed-red) and an inversion symmetry breaking WSM (solid black).  Schematic plots of the two band structures are given as insets.
}
\end{figure}

So far, we have restricted our attention to a single Dirac cone.  This effectively describes DSMs which are inversion- and time-reversal-symmetric.  Here a simple degeneracy factor can be included in the formulas to account for the number of cones (for example, spin and valley degeneracy).  If either symmetry is broken, a WSM is obtained\cite{Wan:2011}.  This leads to a pair of Weyl nodes which have opposite chirality ($\chi=\pm$) and are described by the low-energy Hamiltonian\cite{Vazifeh:2013,Chang:2015}
\begin{align}\label{Ham-Weyl}
\hat{H}=\chi(\hbar v_F\bm{\sigma}\cdot (\bm{q}-\chi\bm{Q})+\mathcal{I}Q_0),
\end{align}
where $\bm{\sigma}$ is a 3D vector of Pauli matrices and $\mathcal{I}$ is the identity matrix.  When $\bm{Q}=\bm{0}$ and $Q_0=0$, this is simply the single-cone Hamiltonian for DSMs which has been used throughout this paper. A finite $\bm{Q}$ corresponds to broken time-reversal symmetry while a nonzero $Q_0$ results from broken inversion symmetry.  For $\bm{Q}\neq \bm{0}$ and $Q_0=0$, the two Weyl nodes have the same energy but are separated in momentum space (see upper-left inset of Fig.~\ref{fig:DW}); while, for finite $Q_0$ and zero $\bm{Q}$, the two nodes sit at the same momentum with one shifted up in energy by $Q_0$ and the other down by the same amount (lower-right inset of Fig.~\ref{fig:DW}). 

We can generalize our results to systems described by Eqn.~\eqref{Ham-Weyl} by defining $\bm{k}=\chi\bm{q}-\bm{Q}$, which effectively eliminates $\bm{Q}$.  The conductivity of a single Weyl node of chirality $\chi$ is then given by Eqn.~\eqref{sigstart-app} with the substitution $\mu\rightarrow\mu-\chi Q_0$ in the Fermi functions. Therefore, all the results presented thus far for a single Dirac cone account for broken inversion and time-reversal symmetry by summing up the single-node results with $\mu\rightarrow\mu-Q_0$ and $\mu\rightarrow\mu+Q_0$.  It is immediately clear that the optical and dc transport properties are unaffected by broken time-reversal symmetry in zero external magnetic field. 
It should be noted that a shift by a finite Q could in principle modify the
impurity scattering itself, an effect that we have not accounted for in this
work.
For noncentrosymmetric WSMs (finite $Q_0$), two steps will be seen in the optical conductivity, one at $\Omega=|2\mu-2Q_0|$ and the other at $|2\mu+2Q_0|$.  For $T=0$, $\Gamma=0$,
\begin{align}\label{DW-IB}
\sigma_{xx}^{\rm IB}(\Omega)&=\frac{e^2}{24\pi\hbar^3v_F}\Omega[\Theta(\Omega-2|\mu_+|)+\Theta(\Omega-2|\mu_-|)],
\end{align}
and
\begin{align}\label{DW-dc}
\sigma_{xx}^{\rm D}(\Omega)&=\frac{e^2}{6\pi\hbar^3v_F}\delta(\Omega)[\mu_+^2+\mu_-^2],
\end{align}
give the interband and intraband contributions to the optical conductivity, respectively, where $\mu_\pm\equiv\mu\pm Q_0$.  This is written here for only one pair of Weyl nodes.  To include more, one multiplies this result by the number of Weyl pairs.  A schematic plot of the interband conductivity is given in Fig.~\ref{fig:DW}.

A number of remarkable features result for the case of the noncentrosymmetric WSM.  First, the Drude conductivity is always finite.  Even at $\mu=0$ (charge neutrality) the Drude weight is given by $2\sigma_0 Q_0^2$.  Secondly, for finite $\mu$, the linear frequency background at higher energy (above the second step in absorption) should have a slope of twice that of the linear background arising out of the first step.  Finally, at $\mu=0$, the 3D DSM and the WSM with finite $\bm{Q}$ will display a single linear background all the way to low frequency passing through the origin.  For the noncentrosymmetric case with $\mu=0$, the conductivity will show zero interband absorption below $\Omega=2|Q_0|$ and a single step at $\Omega=2|Q_0|$ to a linear background.  Clearly, the signature of the noncentrosymmetric WSM will be quite different from the others.  While different in the details, a similar type of behaviour has been discussed for the dynamical conductivity of AA-stacked bilayer graphene\cite{Tabert:2012} which shows energy-shifted Dirac cones.

In the model Hamiltonian of Eqn.~\eqref{Ham-Weyl}, time-reversal symmetry is broken but no external magnetic field is involved.  The application of a magnetic field ($B$) not directly perpendicular to the electric field can pump charge from one Weyl node to the other, an effect known as the chiral anomaly, which we do not consider here.  The effect of such charge pumping on the optical conductivity of a WSM has been presented in Ref.~\cite{Ashby:2014}, where a two-step conductivity is also predicted for finite $B$.  Our results in Fig.~\ref{fig:DW}, pertain to $B=0$.

Other transport properties discussed in this manuscript should also display differences between the noncentrosymmetric WSM and the time-reversal symmetry breaking WSM in a like manner to Fig.~\ref{fig:DW} and Eqns.~\eqref{DW-IB} and ~\eqref{DW-dc}.  That is, the noncentrosymmetric WSM will be a superposition of two versions of the DSM results from previous sections, one for $\mu_+$ and one for $\mu_-$.  Specifically, consider the dc limit of the conductivity given in Eqn.~\eqref{dc-T0} at $T=0$.  Its generalization to finite $Q_0$ for the constant scattering rate (model I) is
\begin{align}
\sigma_{\rm dc}=\sigma_0\left(\frac{\mu_+^2}{\Gamma_0}+\frac{\mu_-^2}{\Gamma_0}+6\Gamma_0\right).
\end{align}
At charge neutrality, $\mu_\pm=\pm Q_0$, and
\begin{align}
\sigma_{\rm dc}=2\sigma_0\left(\frac{Q_0^2}{\Gamma_0}+3\Gamma_0\right),
\end{align}
which is now dominated by the $Q_0$ term in the clean limit ($\Gamma_0\rightarrow 0$).  The ratio of the dc conductivity for time-reversal symmetry breaking ($Q_0\neq 0$) to its $Q_0=0$ value is
\begin{align}
\frac{\sigma_{\rm dc}(Q_0\neq 0)}{\sigma_{\rm dc}(Q_0=0)}=1+\frac{1}{3}\left(\frac{Q_0}{\Gamma_0}\right)^2,
\end{align}
which increases as the square of $Q_0/\Gamma_0$.  Previously, we saw that the chemical potential drops out of the Lorenz number [see Eqn.~\eqref{L-const-T0}] so $L$ will remain unchanged by a shifting of the Dirac cones by $\pm Q_0$ in energy.  At finite temperature, a correction can arise.  As another example, the thermopower [Eqn.~\eqref{S-const-app}] will not change at charge neutrality as the particle and hole contributions will cancel. 

\section{Including a Massless Gap}

The optical conductivity data of Chen \emph{et al.}\cite{Chen:2015} on the semimetal ZrTe$_5$ shows that the dynamical conductivity is nearly linear in photon energy over a range of 150 meV.  In addition, it extrapolates to a small but finite intercept on the Re$\sigma_{xx}(\Omega)$ axis at $T=8$ K.  This was taken as a sign of Dirac semimetallic behaviour.  A similar result was found in pyrochlore Eu$_2$Ir$_2$O$_7$ by Sushkov \emph{et al.}\cite{Sushkov:2015}; although, in this case, the range over which the conductivity displays linearity is only $\sim 10$ meV.  Here, the linear data extrapolates to $\sim 0$ as the photon energy tends to zero.  In the work of Timusk \emph{et al.}\cite{Timusk:2013}, data on the optical conductivity of several quasicrystals showed linearity over a much larger energy range of $\sim 600-1000$ meV.  For Al$_{63.5}$Cu$_{24.5}$Fe$_{12}$ and Al$_{75.5}$Mn$_{20.5}$Si$_{10.1}$, the data also extrapolates to finite intercept as $\Omega\rightarrow 0$.  For Al$_{70}$Pd$_{20}$Re$_{10}$, the extrapolated conductivity crosses the $\Omega$ axis at $\sim 200$ meV.  For Al$_2$Ru, these authors present curves which show two separate regions of linear behaviour; one extrapolates to zero at $\sim 620$ meV while the other crosses the energy axis at $\sim 50$ meV.  Recently, Xu \emph{et al.}\cite{Xu:2015} measured the optical response of the WSM TaAs and found a linear dependence below $\sim 30$ meV (which extrapolates to zero) followed by a Drude below $\sim 10$ meV.  They also found a second linear region between $\sim 30$ meV and 125 meV with a much reduced slope (factor of $\sim 14$) as compared with the first such region.  This line extrapolates to a finite vertical intercept.  There is also an experiment\cite{Orlita:2014} on 3D zinc-blende Hg$_{1-x}$Cd$_x$Te where a linear interband conductivity is seen in zero magnetic field, from $\sim 50-350$ meV and extrapolates to cut the $\Omega$-axis at positive photon energy.  This material has been described as a system with massless fermions with a low-energy bandstructure which has Dirac cones and a flat band at the Dirac point.  Contributions to the interband conductivity arise from both transitions from the flat band to the Dirac cone and between the cones\cite{Orlita:2014,Malcolm:2015}, analogous to what is studied here, and theory predicts a linear conductivity\cite{Orlita:2014,Pavlovic:2015}.  This system can be understood as a superposition between a pseudospin $1/2$ and pseudospin 1 3D DSM\cite{Malcolm:2015, Illes:2015} and, hence, it fits with our discussion.  Overall, it is clear that the optical conductivity of a number of materials has been measured and the unusual linear conductivity suggesting 3D Dirac cones has been seen although some details differ. 

\begin{figure}[h!]
\begin{center}
\includegraphics[width=1.0\linewidth]{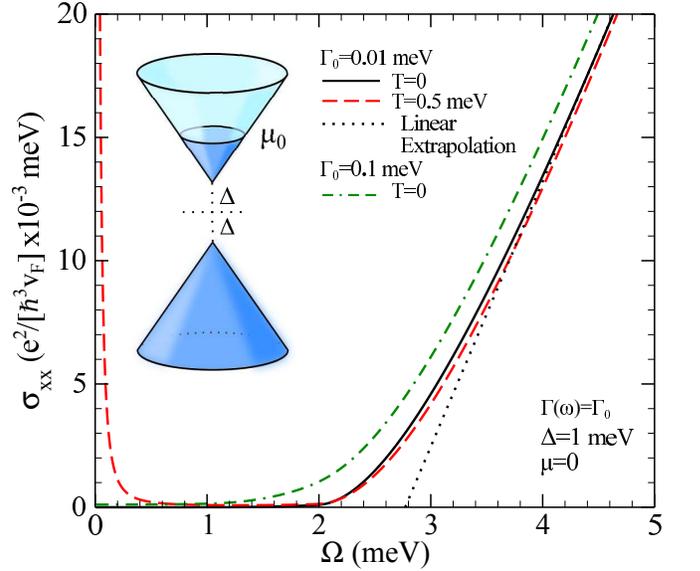}
\end{center}
\caption{\label{fig:Delta-T}(Color online) Optical conductivity of the gapped model. Unlike before, the interband background extrapolates to a negative $y$-intercept. Inset: A 2D plot of the energy dispersion is shown where dark blue colouring signifies occupied states. 
}
\end{figure}

In the simplest theory of a DSM presented so far, the extrapolation of the linear interband background goes through zero (in the clean limit) as it is simply proportional to $\Omega$.  This observation poses an important question: how can our model be modified to give a finite $\Omega$ intercept as seen in a number of experiments?  
The introduction of a Dirac mass $\Delta$ into the electronic
dispersions to get $\epsilon_k=\pm\sqrt{\Delta^2+(\hbar v_F k)^2}$ does
not provide such a shift but rather, at zero temperature, introduces a cut off
in Eq.~(21) for the interband optical conductivity of $2\Delta$. The extrapolation of
$\sigma^{IB}_{xx}(\Omega)= \frac{e^2}{24\pi\hbar^3 v_F}\theta(\Omega-2\Delta)\Omega$
to $\Omega=0$ still goes through the origin. However, as was already noted
by Timusk \emph{et al.}\cite{Timusk:2013}, $\sigma^{IB}_{xx}(\Omega)$ can cut the horizontal
axis at finite photon frequency if the two Dirac cones are pushed up and down
by a gap $\Delta$ as illustrated by the inset in Fig.~\ref{fig:Delta-T}. Such a band structure, which
goes beyond a standard Dirac fermion model, leads naturally to an interband
optical characteristic which goes through zero at finite photon energy $\Omega$,
while massive Dirac cones do not.
A similar model was introduced and studied in relation to 2D systems (namely graphene) in the work of Benfatto \emph{et al.}\cite{Benfatto:2008,Cappelluti:2014}.  In their work, the energy of the Dirac quasiparticles is modified from the usual $\varepsilon_k=\pm\hbar v_Fk$ to $\pm(\hbar v_Fk+\Delta)$.  The authors argue that the
origin of $\Delta$ can be found in the consideration of self-energy effects.
The possible microscopic origin of such a spectrum was elaborated upon
in Ref.~\cite{Cappelluti:2014}. 
The interband conductivity becomes
\begin{align}\label{D-IB-2D}
\sigma_{\rm IB}^{\rm 2D}(\Omega)=\frac{e^2}{4\hbar}\left(1-\frac{2\Delta}{\Omega}\right)\Theta(\Omega-{\rm max}[2\Delta,2\mu])
\end{align}
at $T=0$.  Recently, Morimoto \emph{et al.}\cite{Morimoto:2015} presented similar results for a Mott-Weyl insulator and obtained a 3D version of Eqn.~\eqref{D-IB-2D} $\sigma_{\rm IB}^{\rm 3D}(\Omega)\propto\frac{(\Omega-U)^2}{\Omega}\Theta(\Omega-U)$, where $U$ is a Hubbard potential which plays the role of a gap.  

We can obtain similar results in a 3D band structure model with quasiparticle
energy modified from our previously used $\varepsilon_k=\pm\hbar v_Fk$ to 
$\pm(\hbar v_Fk+\Delta)$.  The interband conductivity at finite temperature for
such a phenomenological band structure model, whatever its origin, is:
\begin{align}
\sigma_{xx}^{\rm IB}(\Omega)&=\frac{e^2}{3\pi\hbar^3v_F}\int_{-\infty}^\infty d\omega\frac{f(\omega)-f(\omega+\Omega)}{\Omega}\int_0^\infty d\varepsilon \varepsilon^2\notag\\
&\times[\delta(\omega-\varepsilon-\Delta)\delta(\omega+\Omega+\varepsilon+\Delta)\notag\\
&+\delta(\omega+\varepsilon+\Delta)\delta(\omega+\Omega-\varepsilon-\Delta)],
\end{align}
which can be reduced to
\begin{align}
\sigma_{xx}^{\rm IB}(\Omega)&=\frac{e^2}{3\pi\hbar^3v_F}\int_{-\infty}^\infty d\omega\frac{f(-\Omega/2)-f(\Omega/2)}{\Omega}\delta(2\omega+\Omega)\notag\\
&\times[(\omega-\Delta)^2\Theta(\omega-\Delta)+(\omega+\Delta)^2\Theta(-\omega-\Delta)].
\end{align}
After manipulation, this reads
\begin{align}
\sigma_{xx}^{\rm IB}(\Omega)&=\frac{e^2}{3\pi\hbar^3v_F}\frac{{\rm sinh}(\beta\Omega/2)}{{\rm cosh}(\beta\mu)+{\rm cosh}(\beta\Omega/2)}\notag\\
&\times\int_{\Delta}^\infty d\omega\frac{(\omega-\Delta)^2}{\Omega}[\delta(2\omega+\Omega)+\delta(-2\omega+\Omega)].
\end{align}
The first $\delta$-function will always give zero in the range $\omega\in [\Delta,\infty)$ leaving
\begin{align}\label{D-IB-T}
\sigma_{xx}^{\rm IB}(\Omega)&=\frac{e^2}{24\pi\hbar^3v_F}\frac{{\rm sinh}(\beta\Omega/2)}{{\rm cosh}(\beta\mu)+{\rm cosh}(\beta\Omega/2)}\notag\\
&\times\frac{(\Omega-\Omega_c)^2}{\Omega}\Theta(\Omega-\Omega_c).
\end{align}
At zero temperature, the thermal factor reduces to the Heaviside step function $\Theta(\Omega-2\mu)$ and we obtain:
\begin{align}\label{D-IB-T0}
\sigma_{xx}^{\rm IB}(\Omega)&=\frac{e^2}{24\pi\hbar^3v_F}\frac{(\Omega-2\Delta)^2}{\Omega}\Theta(\Omega-{\rm max}[2\mu,2\Delta]).
\end{align}

In Fig.~\ref{fig:Delta-T}, we provide numerical results for $\mu=0$ which confirm our analytic formulas derived in the clean limit.  For this figure, however, we return to Eqn.~\eqref{sigstart-app} and include a gap in the dispersion curves through the substitution shown in Eqn.~\eqref{spec-D-app}. The clean limit is not taken so that both contributions from the $\Gamma_0$ and $1/\Gamma_0$ terms are retained. We have included a small constant residual scattering rate $\Gamma(\omega)=\Gamma_0=0.01$ meV, and used $\Delta=1$ meV so that the optical gap is $\Omega_c=2$ meV.  Two values of temperature are considered: $T=0$ (solid black) and $T=0.5$ meV (dashed red).  We also show $T=0$ for $\Gamma_0=0.1$ meV (double-dash-dotted green).  For $T=0$, the solid black curve onsets at 2 meV and then rises according to Eqn.~\eqref{D-IB-T0} before displaying quasilinear behaviour which we have indicated by a dotted line and labelled linear extrapolation.  This line extrapolates to $\Omega\sim 2.8$ meV.  It should be noted that this is not the linear behaviour expected as $\Omega\rightarrow\infty$.  In that limit, the factor $(\Omega-\Omega_c)^2/\Omega$ of Eqn.~\eqref{D-IB-T0} goes like $\Omega-2\Omega_c+\Omega_c^2/\Omega$ which is approximately $\Omega-2\Omega_c$.  This would display an $\Omega$ intercept of $2\Omega_c=4$ meV in our example.  However, in the range of frequency shown here, the quasilinear dependence extrapolates to a lower value $\sim 2.8$ meV.  Our results agree well with the experimental data of Timusk \emph{et al.}\cite{Timusk:2013} on the quasicrystal Al$_2$Ru (higher energy component) with an optical gap of $\approx 600$ meV and with Al$_{70}$Pd$_{20}$Re$_{10}$ where $\Omega_c\approx 200$ meV.  Comparison with the data of Sushkov \emph{et al.}\cite{Sushkov:2015} on pyrochlore Eu$_2$Ir$_2$O$_7$ indicates a negligible gap.  Comparison with Chen \emph{et al.}\cite{Chen:2015} on the semimetal ZrTe$_5$ indicates a small negative gap. Conversely, Xu \emph{et al.}\cite{Xu:2015} suggest that the second linear region observed in the conductivity of TaAs (which does not extrapolate to zero) can be understood as having a significant component coming from trivial bands rather than Weyl points. 

We can also infer from Fig.~\ref{fig:Delta-T}, that at finite $T<\Delta$ (dashed red for $T=\Delta/2$), temperature has little effect on the interband conductivity while it introduces a Drude.  Finally, the double-dash-dotted green curve shows that increasing the scattering rate introduces broadening which fills in the region below the optical gap at $\Omega=2$ meV.  

\section{Discussion and Conclusions}

In this manuscript, we calculate the dc conductivity about a single Dirac or Weyl node within the Kubo formulation.  Particular attention is paid to the approach towards charge neutrality, either at zero temperature in the limit of small chemical potential ($\mu\rightarrow 0$), or at $\mu=0$ for $T\rightarrow 0$.  Three models for the residual (impurity) scattering rate are considered.  Namely, a constant (model I), weak scattering in the Born approximation which gives a scattering rate $\Gamma(\omega)$ proportional to the density of states $N(\omega)\propto\omega^2$ (model II), and long-range Coulomb charged impurities with the number of scattering centres equal to the hole dopping (model III).  These model were considered in the work of Lundgren \emph{et al.}\cite{Lundgren:2014} on dc transport in Weyl and Dirac semimetals with which we wish to compare.  These authors used the Boltzmann equation formulation while we proceed from the Kubo formula.  This provides a correction term which can be of significant importance particularly when $\mu$ is not much greater than the residual scattering rate or temperature.  

The dc conductivity at $\mu=T=0$ [$\sigma_{\rm dc}(\mu=0,T=0)$] depends on the scattering model employed and is not universal (independent of the scattering rate) as it is in graphene.  For example, in model I, it is given by $\sigma_{\rm dc}(\mu,T)=\sigma_0\left[3\Gamma_0+(1/\Gamma_0)\left(\mu^2+\frac{1}{3}\pi^2T^2\right)\right]$ where $\Gamma_0$ is the constant scattering rate.  Therefore, $\sigma_{\rm dc}(\mu=0,T=0)=3\Gamma_0\sigma_0$ with $\sigma_0=e^2/(12\pi^2\hbar^3 v_F)$.  This linearly depends on $\Gamma_0$.  In the Boltzmann approach, the term linear in $\Gamma_0$ is not there and thus $\sigma_{\rm dc}(\mu=0,T=0)=0$ instead.  We also note that in the approach to $(\mu,T)=(0,0)$, the value of $\sigma_{\rm dc}$ is quadratic in $\mu$ and $T$ for both formulations.  This is not the case in model II.  The Kubo formula predicts the same quadratic dependence on $\mu$ and $T$ while the Boltzmann equation gives a constant conductivity with no dependence on $\mu$ or $T$.

When considering the temperature dependence of the dc conductivity, we find it important to account for the $T$ dependence of the chemical potential.  This arises from the energy dependence of the density of states which is quadratic in $\omega$.  These effects are also important for other transport coefficients, such as the thermal conductivity, thermopower (Seebeck coefficient) and Lorenz number (Wiedemann-Franz law).  A minimum is obtained in $\sigma_{\rm dc}$ as a function of $T$ which is traced to the $T$ dependence of $\mu$.  The predicted thermopower shows a much sharper peak at low $T$ and decays much more rapidly as temperature increases when this dependence is included in $\mu$.  It also leads to a faster rise in the Lorenz number ($L$) with $T$.  For all three scattering models, the Wiedemann-Franz law is obeyed at $T=0$ (i.e. $L_0=\pi^2k_B^2/(3e^2)$ where $k_B$ is the Boltzmann constant and $e$ the electron charge).  It rises with $T$ and comes to a high temperature plateau which is $L=4.2L_0$ for models I and II and is $L=9.49L_0$ for the Coulomb scattering model III.  This offers the possibility of differentiating between the models.  In this regard, we have also found that, for weak scattering (model II), the Boltzmann and Kubo approaches predict a quite different high temperature saturated value.  It is $4.2L_0$ in our work whereas keeping only the Boltzmann-like term gives $L_0$ (i.e. there is no $T$ dependence of $L$).

For the constant scattering rate model, we find it possible and instructive to derive simple analytic formulas for the ac conductivity in the clean limit ($\Gamma_0\rightarrow 0$).  The interband transitions at $T=0$ provide a linear-in-$\Omega$ background which extrapolates to zero as $\Omega\rightarrow 0$.  However, the interband response is cut off below $\Omega=2\mu$.  As $T$ increases, for $\mu=0$, the interband background depletes at small $\Omega$ over a range of photon energies of order $\gtrsim 10$ times the temperature.  For finite $\mu$, the jump at $\Omega=2\mu$ becomes smeared and for $T$ of order $\mu$, no identifiable signature of a sharp step remains.  As the interband background becomes depleted by increasing temperature, the Drude contribution is increased and we find a sum rule on the optical spectral weight.  Specifically, the optical spectral weight in the Drude increases as the square of the chemical potential (i.e. $\propto \mu(T)^2$) as well as with the square of temperature ($\sim(\pi^2/3)T^2$).  In general, $\mu(T)$ decreases with increasing $T$ and, in some circumstances, this can decrease the Drude spectral weight by more than the increase from the $(\pi^2/3)T^2$ term.  This leads to a net decreases from its $T=0$ value.  At higher $T$, the Drude weight always equals $\sigma_0\pi^2T^2/3$.

For model I, it is also possible to obtain a simple expression for the ac conductivity at $T=0$ for any value of $\Gamma_0$.  The Drude conductivity is found to be a Lorentzian with an optical scattering width of $2\Gamma_0$ which is multiplied by the sum of three terms.  The first gives the usual Drude Lorentzian of metal theory.  The second is proportional to $(\mu/\Gamma_0)^2$ which does not modify the form of the Lorentzian as a function of $\Omega$ but simply changes the spectral weight.  The final term goes like $[\Omega/(2\Gamma_0)]^2$ and this does modify the shape of the Lorentzian.  In fact, for $\Omega\rightarrow \infty$, it no longer goes to zero but rather becomes a constant equal to $4\Gamma_0\sigma_0/3$.  As with finite $T$, including a nonzero $\Gamma_0$ smears out the interband jump at $\Omega=2\mu$ expected in the clean limit.  It also adds spectral weight in the interband transitions below $2\mu$ and depletes the region above.  For small $\mu/\Gamma_0$, the Drude and interband regions overlap significantly.  The distinction between them becomes clearer as $\mu/\Gamma_0$ increases. This is also the case for model II; however, no simple analytic expressions exist and so we must proceed numerically.  For long-range Coulomb scattering, we find distinct structures appear in $\sigma_{xx}(\Omega)$ due to the complex variation of $\Gamma(\omega)$ vs. $\omega$.  This could provide information on such complicated scattering models.

In the clean limit, an isolated Dirac or Weyl node leads to an interband background which is linear in $\Omega$ and extrapolates to the origin as $\Omega\rightarrow 0$.  Many experiments on Dirac and Weyl materials as well as quasicrystals have found linear interband regions which are taken as signatures of Dirac physics.  However, the data often does not extrapolate to zero as $\Omega\rightarrow 0$.  Instead, it can cross the $y$ axis with a positive intercept, or it can cut the photon-energy axis at $\Omega>0$.  While we have seen that impurity scattering can modify the linear background and provide a positive intercept on the vertical axis, the observation of linear experimental data that intercepts the photon-energy axis at $\Omega>0$ cannot easily be explained.  With this in view, we have also studied a model where a massless gap is included which could provide a more robust explanation of some of the available data.  In 2D, an analogous model was used by Benfatto \emph{et al.}\cite{Benfatto:2008,Cappelluti:2014} to explain the optical data in graphene obtained by Dawlaty \emph{et al.}\cite{Dawlaty:2008}.  In this model, the conduction-band Dirac cone is shifted vertically upward by a gap $\Delta$ while the valence cone is shifted downward by the same amount.  More recently, Morimoto \emph{et al.}\cite{Morimoto:2015} have studied the Mott-Weyl insulator in which the valence and conduction bands are displaced from each other by a vertical gap due to the correlations with the Hubbard $U$ which provides the gap.  Consequently, we have generalized our work to include such a gap. The energy spectrum is thus replaced by $\varepsilon_k=\pm(\hbar v_Fk+\Delta)$ as opposed to the usual $\varepsilon_k=\pm\hbar v_Fk$ dispersion of a 3D Dirac material.  At zero temperature, the $\Omega$ dependence of the interband conductivity calculated in the clean limit is modified to $\sim(\Omega-2\Delta)^2/\Omega$.  When the optical gap $\Omega_c\equiv 2\Delta$ is set to zero, we retain our previous result.  This model has a built in displacement of $\Omega_c$ along the photon-energy axis and is quasilinear at small $\Omega$ before becoming linear when $\Omega\gg\Omega_c$.

Finally, we have discussed predicted differences between the two types of WSMs: those due to time-reversal symmetry breaking which separates the Dirac cones in momentum versus noncentrosymmetric WSMs which have Dirac cones at the same position in momentum but shifted in energy relative to each other.  For the former case, the results discussed here will be indistinguishable from the 3D DSM situation.  However, for the latter system, we predict two steps in the $B=0$ absorptive dynamical conductivity at finite charge doping, each followed by a linearly increasing background with a slope ratio between them of two (for simple Dirac cones).  Unlike the previous cases, there will be a Drude absorption at all dopings, even at charge neutrality where $\mu=0$.  Moreover, at $\mu=0$, there will still be one step in the interband background.  This result is understood as a superposition of two Dirac cones, each with a different doping $\mu_\pm$.  Other properties discussed here (such as the Seebeck coefficient, thermal conductivity and Lorenz number) can also show differences for this type of WSM via the signature of the displacement in energy of the split cones.

The study of the transport and optical properties of Dirac-Weyl semimetals is
still in its infancy and systematic data involving samples with 
well-characterized and controlled defects is largely still lacking. 
Our calculations serve to illustrate that the temperature and chemical 
potential variation of the DC transport coefficients can provide information 
on aspects of the possible microscopic residual scattering models. Studies of 
the intraband and interband AC conductivity can provide additional information.

\begin{acknowledgments}
This work has been supported by the Natural Sciences and Engineering Research Council (NSERC) (Canada) and, in part, by the Canadian Institute for Advanced Research (Canada).
\end{acknowledgments}

\appendix
\section{}\label{app:A}

In the one-loop approximation, the real part of the finite frequency Kubo formula is
\begin{align}\label{Kubo-app}
&\sigma_{\alpha\beta}(\Omega)=\frac{e^2\pi}{\Omega}\int_{-\infty}^\infty d\omega[f(\omega)-f(\omega+\Omega)]\notag\\
&\times\int\frac{d^3k}{(2\pi)^3}{\rm Tr}[\hat{v}_\alpha\hat{\mathcal{A}}(\bm{k},\omega)\hat{v}_\beta\hat{\mathcal{A}}(\bm{k},\omega+\Omega)],
\end{align}
where $f(\omega)=[{\rm exp}([\omega-\mu]/T)+1]^{-1}$ is the Fermi function.  The velocity operator is related to the Hamiltonian ($\hat{H}$) through the relation $\hat{v}_\alpha=(1/\hbar)(\partial\hat{H}/\partial k_\alpha)=v_F\hat{\sigma}_\alpha$.  The spectral functions $\hat{\mathcal{A}}$ are related to the Green's function through
\begin{align}
\hat{\mathcal{G}}(\bm{k},z)=\int_{-\infty}^\infty\frac{\hat{\mathcal{A}}(\bm{k},\omega)}{z-\omega}d\omega,
\end{align}
which is, in turn, defined by
\begin{align}
\hat{\mathcal{G}}^{-1}(\bm{k},z)=\hat{I}z-\hat{H},
\end{align}
where $\hat{I}$ is the identity matrix of dim($\hat{H}$).  The longitudinal conductivity ($\alpha=\beta=x$) is then\cite{Ashby:2014}
\begin{align}\label{sigstart-app}
&\sigma_{xx}(\Omega)=\frac{e^2}{6\pi\hbar^3 v_F}\int_{-\infty}^\infty d\omega\frac{f(\omega)-f(\omega+\Omega)}{\Omega}\notag\\
&\times\int_0^\infty d\varepsilon\varepsilon^2[\hat{\mathcal{A}}_+\hat{\mathcal{A}}_+^\prime+\hat{\mathcal{A}}_-\hat{\mathcal{A}}_-^\prime+2(\hat{\mathcal{A}}_+\hat{\mathcal{A}}_-^\prime+\hat{\mathcal{A}}_-\hat{\mathcal{A}}_+^\prime)],
\end{align}
where $\hat{\mathcal{A}}_\pm\equiv\hat{\mathcal{A}}(\pm\varepsilon,\omega)$ and $\hat{\mathcal{A}}_\pm^\prime\equiv\hat{\mathcal{A}}(\pm\varepsilon,\omega+\Omega)$ with
\begin{align}
\hat{\mathcal{A}}(\pm\varepsilon,\omega)=\frac{1}{\pi}\frac{-{\rm Im}\Sigma(\omega)}{(\omega-{\rm Re}\Sigma(\omega)\mp\varepsilon)^2+{\rm Im}\Sigma(\omega)^2},
\end{align}
where $\Sigma(\omega)$ is the self-energy which contains information on the form of the impurity potential.  Performing the $d\varepsilon$ integral, the intraband component becomes
\begin{align}
\int_0^\infty&\varepsilon^2d\varepsilon[\hat{\mathcal{A}}_+\hat{\mathcal{A}}_+^\prime+\hat{\mathcal{A}}_-\hat{\mathcal{A}}_-^\prime]\notag\\
&=\frac{1}{\pi}\frac{\Gamma^2\Gamma^\prime+\Gamma^\prime(\Sigma_1-\omega)^2+\Gamma({\Gamma^{\prime}}^2+(\omega-\Sigma_1^\prime+\Omega)^2)}{(\Gamma+\Gamma^\prime)^2+(\Sigma_1-\Sigma_1^\prime+\Omega)^2},
\end{align}
where $\Gamma\equiv{\rm Im}\Sigma(\omega)$, $\Gamma^\prime\equiv{\rm Im}\Sigma(\omega+\Omega)$, $\Sigma_1\equiv{\rm Re}\Sigma(\omega)$, and $\Sigma_1^\prime\equiv{\rm Re}\Sigma(\omega+\Omega)$.  Likewise, the interband contribution is
\begin{align}
\int_0^\infty&2\varepsilon^2d\varepsilon [\hat{\mathcal{A}}_+\hat{\mathcal{A}}_-^\prime+\hat{\mathcal{A}}_-\hat{\mathcal{A}}_+^\prime]\notag\\
&=\frac{2}{\pi}\frac{\Gamma^2\Gamma^\prime+\Gamma^\prime(\Sigma_1-\omega)^2+\Gamma({\Gamma^{\prime}}^2+(\omega-\Sigma_1^\prime+\Omega)^2)}{(\Gamma+\Gamma^\prime)^2+(\Sigma_1+\Sigma_1^\prime-2\omega-\Omega)^2}.
\end{align}
For simplicity, we take $\Sigma_1=\Sigma_1^\prime=0$.  Therefore, the intraband and interband pieces of the conductivity are
\begin{align}\label{intra-app}
&\sigma_{xx}^{\rm D}(\Omega)=\frac{e^2}{6\pi^2\hbar^3 v_F}\int_{-\infty}^\infty d\omega\frac{f(\omega)-f(\omega+\Omega)}{\Omega}\notag\\
&\times\frac{\Gamma^2\Gamma^\prime+\Gamma^\prime\omega^2+\Gamma{\Gamma^{\prime}}^2+\Gamma(\omega+\Omega)^2}{(\Gamma+\Gamma^\prime)^2+\Omega^2},
\end{align}
and\begin{align}\label{inter-app}
&\sigma_{xx}^{\rm IB}(\Omega)=\frac{e^2}{3\pi^2\hbar^3 v_F}\int_{-\infty}^\infty d\omega\frac{f(\omega)-f(\omega+\Omega)}{\Omega}\notag\\
&\times\frac{\Gamma^2\Gamma^\prime+\Gamma^\prime\omega^2+\Gamma{\Gamma^{\prime}}^2+\Gamma(\omega+\Omega)^2}{(\Gamma+\Gamma^\prime)^2+(2\omega+\Omega)^2},
\end{align}
respectively.  In the limit of small scattering,  these reduce to Eqns.~(8) and (9) of Ref.~\cite{Ashby:2014}.

In the dc limit, $\Omega\rightarrow 0$, and we have
\begin{align}\label{dc-intra-app}
&\sigma_{\rm dc}^{\rm D}=\frac{e^2}{6\pi^2\hbar^3 v_F}\int_{-\infty}^\infty d\omega\left(-\frac{\partial f(\omega)}{\partial\omega}\right)\frac{\omega^2+\Gamma^2}{2\Gamma},
\end{align}
and\begin{align}\label{dc-inter-app}
&\sigma_{\rm dc}^{\rm IB}=\frac{e^2}{6\pi^2\hbar^3 v_F}\int_{-\infty}^\infty d\omega\left(-\frac{\partial f(\omega)}{\partial\omega}\right)\Gamma,
\end{align}
yielding
\begin{align}\label{dc-app}
&\sigma_{\rm dc}=\frac{e^2}{6\pi^2\hbar^3 v_F}\int_{-\infty}^\infty d\omega\left(-\frac{\partial f(\omega)}{\partial\omega}\right)\frac{\omega^2+3\Gamma^2}{2\Gamma}.
\end{align}
At $T=0$, this gives
\begin{align}
&\sigma_{\rm dc}=\frac{e^2}{12\pi^2\hbar^3 v_F}\frac{\mu^2+3\Gamma(\mu)^2}{\Gamma(\mu)},
\end{align}
where we have explicitly restored the functional dependence of $\Gamma(\omega)$.

To consider a gapped model with gap 2$\Delta$, we return to Eqn.~\eqref{sigstart-app} and make the substitution
\begin{align}\label{spec-D-app}
\int_0^\infty\varepsilon^2 d\varepsilon\rightarrow\int_\Delta^\infty(\varepsilon-\Delta)^2 d\varepsilon.
\end{align}

In the dc limit,
\begin{align}\label{dc-D-app}
\int_\Delta^\infty&(\varepsilon-\Delta)^2d\varepsilon[\hat{\mathcal{A}}_+\hat{\mathcal{A}}_+^\prime+\hat{\mathcal{A}}_-\hat{\mathcal{A}}_-^\prime]\notag\\
&=\frac{1}{2\Gamma\pi^2}\bigg[\pi\Delta^2-2\Delta\Gamma+\pi(\Gamma^2+\omega^2)\notag\\
&-(\Gamma^2+(\Delta-\omega)^2){\rm arctan\left(\frac{\Delta-\omega}{\Gamma}\right)}\notag\\
&-\left.(\Gamma^2+(\Delta+\omega)^2){\rm arctan\left(\frac{\Delta+\omega}{\Gamma}\right)}\right],
\end{align}
and
\begin{align}\label{dc-IB-app}
\int_\Delta^\infty&2(\varepsilon-\Delta)^2d\varepsilon [\hat{\mathcal{A}}_+\hat{\mathcal{A}}_-^\prime+\hat{\mathcal{A}}_-\hat{\mathcal{A}}_+^\prime]\notag\\
&=\frac{\Gamma}{2\pi^2\omega(\Gamma^2+\omega^2)}\bigg[2\pi\omega(\Delta^2+\Gamma^2+\omega^2)\notag\\
&-2[\Delta^2\omega+(\Gamma^2+\omega^2)(\omega-2\Delta)]{\rm arctan}\left(\frac{\Delta-\omega}{\Gamma}\right)\notag\\
&-2[\Delta^2\omega+(\Gamma^2+\omega^2)(\omega+2\Delta)]{\rm arctan}\left(\frac{\Delta+\omega}{\Gamma}\right)\notag\\
&+\left.\Gamma(\Gamma^2+\omega^2-\Delta^2){\rm ln}\left(\frac{\Gamma^2+(\Delta+\omega)^2}{\Gamma^2+(\Delta-\omega)^2}\right)\right].
\end{align}

\section{}\label{app:B}

To study the Wiedemann-Franz law requires information about the electrical and thermal conductivities.  The Lorenz number ($L$) is given by the ratio of the thermal ($\kappa$) and dc-electrical ($\sigma_{\rm dc}$) conductivities as
\begin{align}
L=\frac{\kappa}{T\sigma_{\rm dc}},
\end{align}
where $\sigma_{\rm dc}$ is given by Eqn.~\eqref{dc-app}.  The temperature-scaled thermal conductivity is\cite{Lundgren:2014,Sharapov:2003,Sharma:2015}
\begin{align}\label{kap-app}
\frac{\kappa}{T}=\frac{\kappa_{22}}{T}-\frac{e^2\kappa_{12}^2}{\sigma_{\rm dc}},
\end{align}
where, using the Kubo formula\cite{Sharapov:2003},
\begin{align}\label{kap22-app}
\frac{\kappa_{22}}{T}=\frac{1}{6\pi^2\hbar^3 v_F}\int_{-\infty}^\infty d\omega \left(\frac{\omega-\mu}{T}\right)^2\left(-\frac{\partial f(\omega)}{\partial\omega}\right)\frac{\omega^2+3\Gamma^2}{2\Gamma},
\end{align}
and
\begin{align}\label{kapp12-app}
\kappa_{12}=\frac{1}{6\pi^2\hbar^3 v_F}\int_{-\infty}^\infty d\omega\left(\frac{\omega-\mu}{T}\right)\left(-\frac{\partial f(\omega)}{\partial\omega}\right)\frac{\omega^2+3\Gamma^2}{2\Gamma},
\end{align}
which have similar forms to Eqn. (5.8) of Sharapov \emph{et al.}\cite{Sharapov:2003} but applied to our model.

\subsection{$\Gamma(\omega)=\Gamma_0$}

Using a constant scattering rate, the dc electrical conductivity can be written as
\begin{align}\label{dcT-app1}
\frac{\sigma_{\rm dc}}{\sigma_0\Gamma_0}=\left(\frac{T}{\Gamma_0}\right)^2\mathcal{F}_2^{\bar{\mu}}+3\mathcal{F}_0^{\bar{\mu}},
\end{align}
where
\begin{align}
\mathcal{F}_n^{\bar{\mu}}\equiv\int_{-\infty}^\infty\frac{x^ndx}{4{\rm cosh}^2\left(\frac{x-\bar{\mu}}{2}\right)},
\end{align}
$\sigma_0=1/(12\pi^2\hbar^3v_F)$ and $\bar{\mu}\equiv\mu/T$.  Likewise,
\begin{align}
\frac{\kappa_{22}}{T}\frac{e^2}{\sigma_0\Gamma_0}&=\left(\frac{T}{\Gamma_0}\right)^2\left[\mathcal{F}_4^{\bar{\mu}}+\bar{\mu}^2\mathcal{F}_2^{\bar{\mu}}-2\bar{\mu}\mathcal{F}_3^{\bar{\mu}}\right]\notag\\
&+3\left[\mathcal{F}_2^{\bar{\mu}}+\bar{\mu}^2\mathcal{F}_0^{\bar{\mu}}-2\bar{\mu}\mathcal{F}_1^{\bar{\mu}}\right],
\end{align}
and
\begin{align}
\frac{e^2\kappa_{12}^2}{\sigma_{\rm dc}}\frac{e^2}{\sigma_0\Gamma_0}&=\frac{1}{\bar{\sigma}_{\rm dc}}\left\lbrace\left(\frac{T}{\Gamma_0}\right)^2\left(\mathcal{F}_3^{\bar{\mu}}-\bar{\mu}\mathcal{F}_2^{\bar{\mu}}\right)\right.\notag\\
&+\left.3\left(\mathcal{F}_1^{\bar{\mu}}-\bar{\mu}\mathcal{F}_0^{\bar{\mu}}\right)	\right\rbrace^2,
\end{align}
where $\bar{\sigma}_{\rm dc}=\sigma_{\rm dc}/(\sigma_0\Gamma_0)$.  These expressions are further simplified by noting (for $\bar{\mu}\geq 0$)
\begin{align}\label{F-app}
\mathcal{F}_0^{\bar{\mu}}&=1,\notag\\
\mathcal{F}_1^{\bar{\mu}}&=\bar{\mu},\notag\\
\mathcal{F}_2^{\bar{\mu}}&=\bar{\mu}^2+\frac{\pi^2}{3},\\
\mathcal{F}_3^{\bar{\mu}}&=\bar{\mu}\left(\bar{\mu}^2+\pi^2\right),\notag\\
\mathcal{F}_4^{\bar{\mu}}&=\bar{\mu}^4+2\bar{\mu}^2\pi^2+\frac{7\pi^4}{15}.\notag
\end{align}
Using these results, the Lorenz number is found to be
\begin{align}\label{L-M1-appB}
L=L_0&\left\lbrace\frac{3(15\Gamma_0^2+5\mu^2+7\pi^2T^2)}{5(9\Gamma_0^2+3\mu^2+\pi^2T^2)}\right.\notag\\
&-\left.\frac{12\mu^2\pi^2T^2}{(9\Gamma_0^2+3\mu^2+\pi^2T^2)^2}\right\rbrace,
\end{align}
where $L_0=\pi^2/(3e^2)$.

The thermopower is also readily evaluated and is given by
\begin{align}\label{S-const-app}
S&=\frac{e\kappa_{12}}{\sigma_{\rm dc}}\notag\\
&=\frac{1}{e\bar{\sigma}_{\rm dc}}\left[\left(\frac{T}{\Gamma_0}\right)^2\left(\mathcal{F}_3^{\bar{\mu}}-\bar{\mu}\mathcal{F}_2^{\bar{\mu}}\right)\right.\notag\\
&\quad\quad\left.+3(\mathcal{F}_1^{\bar{\mu}}-\bar{\mu}\mathcal{F}_0^{\bar{\mu}})\right]\notag\\
&=\frac{1}{e}\frac{2\pi^2T\mu}{3\mu^2+\pi^2T^2+9\Gamma_0^2}.
\end{align}

\subsection{$\Gamma(\omega)=\Gamma_1\omega^2$}

If we take the energy dependent scattering rate $\Gamma(\omega)=\Gamma_1\omega^2$, the dc electrical conductivity becomes
\begin{align}\label{dcT-app2}
\frac{\Gamma_1\sigma_{\rm dc}}{\sigma_0}=\mathcal{F}_0^{\bar{\mu}}+3(\Gamma_1T)^2\mathcal{F}_2^{\bar{\mu}}.
\end{align}
Similarly,
\begin{align}
\frac{\kappa_{22}}{T}\frac{e^2\Gamma_1}{\sigma_0}=&3(\Gamma_1T)^2[\mathcal{F}_4^{\bar{\mu}}+\bar{\mu}^2\mathcal{F}_2^{\bar{\mu}}-2\bar{\mu}\mathcal{F}_3^{\bar{\mu}}]\notag\\
&+\mathcal{F}_2^{\bar{\mu}}+\bar{\mu}^2\mathcal{F}_0^{\bar{\mu}}-2\bar{\mu}\mathcal{F}_1^{\bar{\mu}},
\end{align}
and
\begin{align}
\frac{e^2\kappa_{12}^2}{\sigma_{\rm dc}}\frac{e^2\Gamma_1}{\sigma_0}&=\frac{1}{\tilde{\sigma}_{\rm dc}}\left\lbrace 3(\Gamma_1T)^2(\mathcal{F}_3^{\bar{\mu}}-\bar{\mu}\mathcal{F}_2^{\bar{\mu}})\right.\notag\\
&+\left. (\mathcal{F}_1^{\bar{\mu}}-\bar{\mu}\mathcal{F}_0^{\bar{\mu}})\right\rbrace^2,
\end{align}
where $\tilde{\sigma}_{\rm dc}=\Gamma_1\sigma_{\rm dc}/\sigma_0$.  This yields
\begin{align}\label{L-M2-appB}
L=L_0&\left\lbrace\frac{5+3\Gamma_1^2(5\mu^2+7\pi^2T^2)}{5(1+\Gamma_1^2[3\mu^2+\pi^2T^2])}\right.\notag\\
&-\left.\frac{12\pi^2\Gamma_1^4\mu^2T^2}{(1+\Gamma_1^2[3\mu^2+\pi^2T^2])^2}\right\rbrace.
\end{align}
The associated thermopower is
\begin{align}
S=\frac{1}{e}\frac{2\pi^2\Gamma_1^2T\mu}{1+\Gamma_1^2(3\mu^2+\pi^2T^2)}.
\end{align}

\bibliographystyle{apsrev4-1}
\bibliography{WeylCond}

\end{document}